\documentclass[aps, prd, floatfix, nofootinbib, superscriptaddress, twocolumn]{revtex4-1}

\usepackage{latexsym}
\usepackage{amsmath}
\usepackage{amssymb}
\usepackage{amsfonts}

\usepackage[mathscr,scaled=1.15]{urwchancal}
\DeclareFontFamily{OT1}{pzc}{}
\DeclareFontShape{OT1}{pzc}{m}{it}%
{<-> s * [1.15] pzcmi7t}{}
\DeclareMathAlphabet{\mathpzc}{OT1}{pzc}{m}{it}

\usepackage{color}

\usepackage{supertabular}
\usepackage{placeins}
\usepackage{epsfig}
\usepackage{graphicx}

\definecolor{purple}{rgb}{0.5,0,0.5}
\definecolor{blue}{rgb}{0.0,0,0.9}
\definecolor{prdblue}{rgb}{0.133,0.118,0.498}
\usepackage[colorlinks=true, pdfstartview=FitV, linkcolor=prdblue, citecolor= prdblue, urlcolor=prdblue]{hyperref}

\begin{document}


\title{Masses of ground-state mesons and baryons, including those with heavy quarks}



\author{Pei-Lin Yin}
\email[]{yinpl@njupt.edu.cn}
\affiliation{College of Science, Nanjing University of Posts and Telecommunications, Nanjing 210023, China}

\author{Chen Chen}
\email[]{chenchen@ift.unesp.br}
\affiliation{Instituto de F\'isica Te\'orica, Universidade Estadual Paulista, Rua Dr.~Bento Teobaldo Ferraz, 271, 01140-070 S\~ao Paulo, SP, Brazil}

\author{Gast\~ao Krein}
\affiliation{Instituto de F\'isica Te\'orica, Universidade Estadual Paulista, Rua Dr.~Bento Teobaldo Ferraz, 271, 01140-070 S\~ao Paulo, SP, Brazil}

\author{Craig D. Roberts}
\email[]{cdroberts@anl.gov}
\affiliation{Physics Division, Argonne National Laboratory, Lemont, Illinois
60439, USA}

\author{Jorge Segovia}
\affiliation{Departamento de Sistemas F\'{\i}sicos, Qu\'{\i}micos y Naturales,
Universidad Pablo de Olavide, E-41013 Sevilla, Spain}


\author{Shu-Sheng Xu}
\affiliation{College of Science, Nanjing University of Posts and Telecommunications, Nanjing 210023, China}



\date{28 February 2019}

\begin{abstract}
Using a confining, symmetry-preserving regularisation of a vector$\times$vector contact interaction, we compute the spectra of ground-state pseudoscalar and vector $(f\bar g)$ mesons, scalar and axial-vector $(fg)$ diquarks, and $J^P=1/2^+, 3/2^+$ $(fgh)$ baryons, where $f,g,h \in \{u,d,s,c,b\}$.  The diquark correlations are essentially dynamical and play a key role in formulating and solving the three-valence-quark baryon problems.  The baryon spectrum obtained from this largely-algebraic approach reproduces the 22 known experimental masses with an accuracy of $2.9(2.4)$\%.  It also possesses the richness of states typical of constituent-quark models, predicting many heavy-quark baryons not yet observed.  This study indicates that diquark correlations are an important component of all baryons; and owing to the dynamical character of the diquarks, it is typically the lightest allowed diquark correlation which defines the most important component of a baryon's Faddeev amplitude.
\end{abstract}

\maketitle


\section{Introduction}
There are five flavours of quarks which live long enough to produce measurable bound states: $f\in \{u,d,s,c,b\}$; and contemporary theory and phenomenological models predict the existence of bound systems with every allowed colour-singlet $(f \bar{f}^\prime)$-combination for mesons and $(f f^\prime f^{\prime\prime})$-combination for baryons.  In fact, the predicted meson and baryon spectra are so rich that, even regarding ground-states in the $J^P$-channels accessible to constituent quark models, many bound-states are ``missing'', \emph{i.e}.\ have not been observed in experiment.  (See, e.g.\ the quark model review in Ref.\,\cite{Tanabashi:2018oca}.)  This challenge has been accepted, with an array of dedicated experiments underway, at facilities worldwide, which seek to detect the missing states \cite{Akhunzyanov:2018lqa, Carman:2018fsn, Thiel:2018eif, Sokhoyan:2018geu, Cole:2018faq, Galanti:2018bnp}.

Regarding computations of such spectra, the numerical simulation of lattice-regularised quantum chromodynamics (lQCD) provides the most direct connection with the standard model of particle physics and many separate efforts are underway.  Some of the successes and challenges are described, \emph{e.g}.\ in Refs.\,\cite{Fodor:2012gf, Briceno:2017max}; and some recent spectrum calculations are reported in Refs.\,\cite{Brown:2014ena, Alexandrou:2017xwd, Bali:2017pdv, Lubicz:2017asp, Mathur:2018epb}.

The Dyson-Schwinger equations (DSEs) \cite{Roberts:1994dr}, a collection of coupled integral equations that provide for a symmetry-preserving treatment of the continuum bound-state problem, have also been widely employed to compute hadron spectra and interactions \cite{Roberts:2015lja, Horn:2016rip, Eichmann:2016yit, Burkert:2017djo}.
In this connection, the last decade has seen marked improvements in both (\emph{i}) understanding the limitations and capacities of the approach and (\emph{ii}) the breadth and quality of the description of hadron properties, including the spectrum of states \cite{Chang:2011ei, Blank:2011ha, Hilger:2014nma, Sanchis-Alepuz:2015qra, Gomez-Rocha:2015qga, Segovia:2015hra, Segovia:2015ufa, Gomez-Rocha:2016cji, Eichmann:2016hgl, Hilger:2017jti, Burkert:2017djo, Chen:2017pse, Chen:2019fzn}.

Most recently, unified predictions for the spectra of mesons and baryons in some of the low-lying flavour-$SU(N_f=5)$ multiplets were delivered \cite{Qin:2018dqp, Qin:2019hgk} using the rainbow-ladder (RL) truncation of the bound-state equations, which is the leading-order in a systematic scheme \cite{Munczek:1994zz, Bender:1996bb, Binosi:2016rxz}.  Whilst the coverage is still not as extensive as that provided by constituent-quark potential models, whose applications are canvassed in, \emph{e.g}.\, Refs.\,\cite{Godfrey:1985xj, Roberts:2007ni, Crede:2013sze, Segovia:2013wma, Giannini:2015zia, Chen:2016spr, Ebert:2017bnn, Yang:2017qan}, the systematic RL studies have the advantage of providing both (\emph{i}) a unified, symmetry-preserving description of mesons and baryons and (\emph{ii}) a traceable connection to quantum chromodynamics (QCD).  In such continuum studies, a next challenge is to proceed beyond the leading-order truncation, an effort which is likely to benefit from the use of high-performance computing.

Herein, on the other hand, we follow a different path and adopt a largely algebraic approach.  Namely, we exploit the fact that the mass of any given hadron is an integrated (long-wavelength) quantity and thus not very sensitive to details of the system's wave function.  This being so, then it ought to be possible to use a vector$\times$vector contact interaction to obtain both sound predictions for the ground-state spectrum of $SU(N_f=5)$ mesons and baryons, and reliable insights into aspects of their structure.  This has proven true for $SU(N_f=3)$ systems \cite{GutierrezGuerrero:2010md,Roberts:2010rn, Roberts:2011cf, Roberts:2011wy, Wilson:2011aa, Chen:2012qr, Chang:2012cc, Chen:2012txa, Wang:2013wk, Segovia:2013rca, Segovia:2013uga, Xu:2015kta, Lu:2017cln, Raya:2018ioy} and for mesons with one or more heavy quarks \cite{Bedolla:2015mpa, Bedolla:2016yxq, Serna:2017nlr, Bedolla:2018ytg}.  Hence, our goal is to minimally refine the vector$\times$vector contact interaction and therewith deliver a unified description of all mesons and baryons in the ground-state multiplets of $SU(N_f=5)$, anticipating that the framework's simplicity will enable insights into features of these systems that are obscured in approaches that rely heavily on computer resources.

We describe our formulation of the contact interaction in Sec.\,\ref{Sec2}, including a discussion of the regularisation procedure.   In addition to the four current-quark masses (isospin symmetry is assumed), four parameters occur in using the interaction to compute meson properties and we also explain how they are determined.
The calculation of meson properties is detailed in Sec.\,\ref{SecMeson}.

In computing baryon spectra, we use a quark+diquark approximation to the three-valence-quark problem and hence require the masses and correlation amplitudes for each diquark that can play a role.  Their calculation is explained in Sec.\,\ref{SecDiquarks}.
The formulation and solution of the baryon problem is discussed in Sec.\,\ref{SecBaryons}.  It includes results for the spectra of ground-state flavour-$SU(5)$ $J^P=1/2^+, 3/2^+$ baryons and their contact-interaction Faddeev amplitudes, which are momentum-independent.

Section~\ref{epilogue} presents a summary and also a perspective on extensions of this study, including new directions.

\section{Contact Interaction}
\label{Sec2}
\subsection{Two-Body Scattering Kernel}
\label{SeccalG}
The key element in analyses of the continuum bound-state problem for hadrons is the quark-quark scattering kernel.  In RL truncation that can be written ($k = p_1-p_1^\prime = p_2^\prime -p_2$):
\begin{subequations}
\label{KDinteraction}
\begin{align}
\mathscr{K}_{\alpha_1\alpha_1',\alpha_2\alpha_2'}  & = {\mathpzc G}_{\mu\nu}(k) [i\gamma_\mu]_{\alpha_1\alpha_1'} [i\gamma_\nu]_{\alpha_2\alpha_2'}\,,\\
 {\mathpzc G}_{\mu\nu}(k)  & = \tilde{\mathpzc G}(k^2) T_{\mu\nu}(k)\,,
\end{align}
\end{subequations}
where $k^2 T_{\mu\nu}(k) = k^2 \delta_{\mu\nu} - k_\mu k_\nu$.
(Our Euclidean metric and Dirac-matrix conventions are specified in Ref.\,\cite{Chen:2012qr}, Appendix~A.)
The defining quantity is $\tilde{\mathpzc G}$; and following two decades of study, much has been learnt about its pointwise behaviour using a combination of continuum and lattice methods in QCD \cite{Binosi:2014aea, Binosi:2016nme, Rodriguez-Quintero:2018wma}.  The qualitative conclusion is that owing to the dynamical generation of a gluon mass-scale in QCD \cite{Bowman:2004jm, Boucaud:2006if, Aguilar:2008xm, Boucaud:2011ug, Aguilar:2012rz, Ayala:2012pb, Strauss:2012dg, Aguilar:2015bud, Binosi:2016xxu, Gao:2017uox}, $\tilde{\mathpzc G}$ saturates at infrared momenta; hence, one may write
\begin{align}
\tilde{\mathpzc G}(k^2) & \stackrel{k^2 \simeq 0}{=} \frac{4\pi \alpha_{IR}}{m_G^2}\,.
\end{align}

In QCD, $m_G \approx 0.5\,$GeV, $\alpha_{\rm IR} \approx \pi$ \cite{Binosi:2016nme, Rodriguez-Quintero:2018wma}.  We keep this value of $m_G$ in developing the contact-interaction for use in RL truncation, but reduce $\alpha_{IR}$ to a parameter.  The latter is necessary because the integrals that appear in contact-interaction bound-state equations require ultraviolet regularisation and this spoils the intimate connection between infrared and ultraviolet scales that is a hallmark of QCD.  In addition, since a contact interaction cannot support relative momentum between bound-state constituents, we simplify the tensor structure in Eqs.\,\eqref{KDinteraction} and define the contact-interaction RL kernel as follows:
\begin{align}
\label{KCI}
\mathscr{K}_{\alpha_1\alpha_1',\alpha_2\alpha_2'}^{\rm CI}  & = \frac{4\pi \alpha_{IR}}{m_G^2}
 [i\gamma_\mu]_{\alpha_1\alpha_1'} [i\gamma_\mu]_{\alpha_2\alpha_2'}\,.
 \end{align}

As just remarked, any use of Eq.\,\eqref{KCI} in a DSE will require imposition of an ultraviolet regularisation scheme, which should be symmetry-preserving; and since the theory is not renormalisable, the associated mass-scales, $\Lambda_{\rm uv}$, will be additional physical parameters.  It is useful to interpret any one of these scales as an upper bound on the momentum domain within which the properties of the associated system are effectively momentum-independent, \emph{e.g}.\ the $\pi$-meson has a larger radius than the $\eta_b$; hence one should expect to use $1/\Lambda_{\rm uv}^\pi > 1/\Lambda_{\rm uv}^{\eta_c}$.  The implications of this approach will subsequently be elucidated.

We will also introduce an infrared regularisation scale, $\Lambda_{\rm ir}$, when defining those integrals that contribute to the bound-state problems \cite{Ebert:1996vx}.  Since chiral symmetry is dynamically broken in our approach, ensuring the absence of infrared divergences, $\Lambda_{\rm ir}$ is not a necessary part of the contact-interaction's definition.  On the other hand, by excising momenta less-than $\Lambda_{\rm ir}$, one achieves a rudimentary expression of confinement via elimination of quark production thresholds \cite{Krein:1990sf, Roberts:2007ji, Horn:2016rip}.  A natural choice for this scale is $\Lambda_{\rm ir} \sim \Lambda_{\rm QCD}$ and we set $\Lambda_{\rm ir} = 0.24\,$GeV.

\subsection{Interaction Scales}
\label{SecIS}
In order to fix the parameters in our implementation of Eq.\,\eqref{KCI} we focus on the masses and leptonic decay constants of the following pseudoscalar mesons: $\pi$, $K$, $\eta_c$, $\eta_b$; whose valence-quark content is $f \bar f$, $f\in \{l=u=d, s, c, b\}$, where isospin symmetry is assumed.

The simplest DSE relevant to the associated bound-state problems is the dressed-quark gap equation.  In RL truncation, using Eq.\,\eqref{KCI}, it takes the following form:
\begin{align}
\label{GapEqn}
S_f^{-1}(p) & = i\gamma\cdot p +m_f
+ \frac{16 \pi}{3} \frac{\alpha_{\rm IR}}{m_G^2}
\int \frac{d^4q}{(2\pi)^4} \gamma_\mu S_f(q) \gamma_\mu\,,
\end{align}
where $m_f$ is the quark's current-mass.  The integral is quadratically divergent; but when it is regularised in a Poincar\'e-invariant manner, the solution is
\begin{equation}
\label{genS}
S_f(p)^{-1} = i \gamma\cdot p + M_f\,,
\end{equation}
where $M_f$ is momentum-independent and determined by
\begin{equation}
M_f = m_f + M_f\frac{4\alpha_{\rm IR}}{3\pi m_G^2} \left[\int_0^\infty \!ds \, s\, \frac{1}{s+M_f^2}\right]_{\rm reg}\,.
\end{equation}

Following Ref.\,\cite{Ebert:1996vx}, we define the integral by writing
\begin{align}
\nonumber
\frac{1}{s+M^2} & = \int_0^\infty d\tau\,{\rm e}^{-\tau (s+M^2)}  \\
&\rightarrow  \int_{\tau_{\rm uv}^2}^{\tau_{\rm ir}^2} d\tau\,{\rm e}^{-\tau (s+M^2)}
\label{RegC}\\
 &=
\frac{{\rm e}^{- (s+M^2)\tau_{\rm uv}^2}-{\rm e}^{-(s+M^2) \tau_{\rm ir}^2}}{s+M^2} \,, \label{ExplicitRS}
\end{align}
where $\tau_{\rm ir,uv}=1/\Lambda_{{\rm ir},{\rm uv}}$ are, respectively, the infrared and ultraviolet regulators described above.  Consequently, the gap equation becomes
\begin{equation}
M_f = m_f + M_f\frac{4\alpha_{\rm IR}}{3\pi m_G^2}\,\,{\cal C}_0^{\rm iu}(M_f^2)\,,
\label{gapactual}
\end{equation}
where
\begin{align}
\nonumber
{\cal C}_0^{\rm iu}(\sigma) &=
\int_0^\infty\! ds \, s \int_{\tau_{\rm uv}^2}^{\tau_{\rm ir}^2} d\tau\,{\rm e}^{-\tau (s+\sigma)}\\
& =
\sigma \big[\Gamma(-1,\sigma \tau_{\rm uv}^2) - \Gamma(-1,\sigma \tau_{\rm ir}^2)\big],
\label{eq:C0}
\end{align}
with $\Gamma(\alpha,y)$ being the incomplete gamma-function.

Using a contact interaction, the Bethe-Salpeter amplitude for a pseudoscalar meson constituted from a valence $f$-quark and valence $g$-antiquark has the following restricted form \cite{GutierrezGuerrero:2010md, Chen:2012txa}:
\begin{align}
\Gamma_{0^-}(Q) = \gamma_5 \left[ i E_{0^-} + \frac{1}{2 M_R}\gamma\cdot Q F_{0^-}\right]\,,
\end{align}
Here, $Q$ is the bound-state's total momentum, $Q^2 = -m_{0^-}^2$, $m_{0^-}$ is the meson's mass; and $M_R= M_f M_{g}/[M_f + M_{g}]$, with $M_{f,g}$ being the relevant dressed-quark masses obtained from the contact-interaction gap equations, described above.

The amplitude is determined by the following equation:
\begin{equation}
\Gamma_{0^-}(Q) =  - \frac{16 \pi}{3} \frac{\alpha_{\rm IR}}{m_G^2}
\int \! \frac{d^4t}{(2\pi)^4} \gamma_\mu S_f(t+Q) \Gamma_{0^-}(Q)S_g(t) \gamma_\mu \,.
\label{LBSEI}
\end{equation}
From here, using the symmetry-preserving regularisation scheme introduced in Refs.\,\cite{GutierrezGuerrero:2010md, Chen:2012txa}, which requires, in the spirit of dimensional regularisation,
\begin{equation}
0 = \int_0^1d\alpha \,
\big[ {\cal C}_0^{\rm iu}(\omega_{fg}(\alpha,Q^2))
%
+ \, {\cal C}^{\rm iu}_1(\omega_{f g}(\alpha,Q^2))\big], \label{avwtiP}
\end{equation}
where ${\cal C}^{\rm iu}_1$ is given in Eqs.\,\eqref{Cndef}, \eqref{C012def} and
\begin{align}
\omega_{f g}(\alpha,Q^2) &= M_f^2 \hat \alpha + \alpha M_{g}^2 + \alpha \hat\alpha Q^2\,,
\label{eq:omega}
\end{align}
$\hat \alpha = 1-\alpha$, one arrives at the following explicit form of the Bethe-Salpeter equation (BSE), Eq.\,\eqref{LBSEI}:
\begin{equation}
\label{bsefinalE}
\left[
\begin{array}{c}
E_{0^-}(Q)\\
F_{0^-}(Q)
\end{array}
\right]
= \frac{4 \alpha_{\rm IR}}{3\pi m_G^2}
\left[
\begin{array}{cc}
{\cal K}_{EE}^{0^-} & {\cal K}_{EF}^{0^-} \\
{\cal K}_{FE}^{0^-} & {\cal K}_{FF}^{0^-}
\end{array}\right]
\left[\begin{array}{c}
E_{0^-}(Q)\\
F_{0^-}(Q)
\end{array}
\right],
\end{equation}
with the matrix elements $\{ {\cal K}_{EE}^{0^-}, {\cal K}_{EF}^{0^-}, {\cal K}_{FE}^{0^-}, {\cal K}_{FF}^{0^-}\}$ defined in Eqs.\,\eqref{fgKernel}.

It is important to note that ${\cal K}_{FE}^{0^-}\neq 0$ when chiral symmetry is dynamically broken; hence, any internally-consistent description of a pseudoscalar meson must retain the state's $F_{0^-}(Q)$ (pseudovector) component.  Models of the Nambu--Jona-Lasinio type that omit this component do not have any connection with an underlying theory whose dynamics is based on vector-boson exchange.  Therefore, they cannot serve as a veracious model of QCD in any energy range.

Eq.\,\eqref{bsefinalE} is an eigenvalue problem that has a solution when $Q^2=-m_{0^-}^2$, at which point the eigenvector is the meson's Bethe-Salpeter amplitude.  Working with the on-shell solution, normalised canonically according to Eqs.\,\eqref{normcan}, \eqref{normcan2}, the pseudoscalar meson's leptonic decay constant is given by:
\begin{align}
f_{0^-} &= \frac{N_c}{4\pi^2}\frac{1}{ M_{R}}\,
\big[ E_{0^-} {\cal K}_{FE}^{0^-} + F_{0^-}{\cal K}_{FF}^{0^-} \big]_{Q^2=-m_{0^-}^2}\,. \label{ffg}
\end{align}

\begin{table}[t]
\caption{\label{Tab:DressedQuarks}
Couplings, ultraviolet cutoffs and current-quark masses that provide a good description of pseudoscalar meson properties, along with the dressed-quark masses and selected pseudoscalar meson properties they produce; all obtained with $m_G=0.5\,$GeV, $\Lambda_{\rm ir} = 0.24\,$GeV.
Empirically, at the level we are working \cite{Tanabashi:2018oca}:
$m_\pi =0.14$, $f_\pi=0.092$; $m_K=0.50$, $f_K=0.11$; $m_{\eta_c} =2.98$, $f_{\eta_c}=0.24$; $m_{\eta_b}=9.40$.
The value of $f_{\eta_b}$ is discussed in connection with Eq.\,\eqref{fetab}.
(Dimensioned quantities in GeV.)}
\begin{center}
\begin{tabular*}
{\hsize}
{
c@{\extracolsep{0ptplus1fil}}|
c@{\extracolsep{0ptplus1fil}}
|c@{\extracolsep{0ptplus1fil}}
c@{\extracolsep{0ptplus1fil}}
|c@{\extracolsep{0ptplus1fil}}
c@{\extracolsep{0ptplus1fil}}
c@{\extracolsep{0ptplus1fil}}}\hline\hline
quark & $\alpha_{\rm IR}/\pi\ $ & $\Lambda_{\rm uv}$ & $m$ &   $M$ &  $m_{0^-}$ & $f_{0^-}$ \\\hline
$l=u/d\ $  & $0.36\phantom{1}$ & $0.91\ $ & $0.007\ $ & 0.37$\ $ & 0.14 & 0.10  \\\hline
$s$  & $0.36\phantom{1}$ & $0.91\ $ & $0.17\phantom{7}\ $ & 0.53$\ $ & 0.50 & 0.11 \\\hline
$c$  & $0.054$ & $1.88\ $ & $1.24\phantom{7}\ $ & 1.60$\ $ & 2.98 & 0.24 \\\hline
$b$  & $0.012$ & $3.50\ $ & $4.66\phantom{7}\ $ & 4.83$\ $ & 9.40 & 0.41
\\\hline\hline
\end{tabular*}
\end{center}
\end{table}

Light-quark systems were analysed in Refs.\,\cite{Roberts:2011cf, Chen:2012txa}, with the results listed in Table~\ref{Tab:DressedQuarks}.  Notably, the fitted value of $m_s/m_l = 24$ is compatible with estimates in QCD \cite{Tanabashi:2018oca}, even though our individual current-masses are too large by a factor of $\lesssim 2$ owing to the contact-interaction's deficiencies in connection with ultraviolet quantities.  Moreover, the result $M_s/M_l =1.4$ for the dressed-quark masses is typical of efficacious DSE studies in the light-quark sector \cite{Maris:1997tm, Shi:2015esa}.

Moving to heavy-quark systems, we allow $\Lambda_{\rm uv}^{0^-}$ to vary with the meson's mass and fix the associated coupling by requiring 
\begin{equation}
\alpha_{\rm IR}(\Lambda_{\rm uv}^{0^-}) [\Lambda_{\rm uv}^{0^-}]^2 \ln\frac{\Lambda_{\rm uv}^{0^-}}{\Lambda_{\rm ir}}
=
\alpha_{\rm IR}(\Lambda_{\rm uv}^{\pi}) [\Lambda_{\rm uv}^{\pi}]^2 \ln\frac{\Lambda_{\rm uv}^{\pi}}{\Lambda_{\rm ir}}\,.
\label{alphaLambda}
\end{equation}
(Similar expedients were adopted in Refs.\,\cite{Bedolla:2015mpa, Serna:2017nlr}.)
This identity serves to limit the number of parameters, so that in fitting the $\eta_{c,b}$ quantities in Table~\ref{Tab:DressedQuarks} we had only two parameters for each case: $m_{c,b}$, $\Lambda_{\rm uv}^{\eta_{c,b}}$.  Regarding the $\eta_b$-meson, a lQCD calculation reports $f_{\eta_b}=0.472(4)$ \cite{McNeile:2012qf}.  However, this is larger than the result for $f_{\Upsilon}=0.459(22)$ \cite{Colquhoun:2015oha}; hence, it is contrary to the experimental pattern: $f_{\pi}<f_{\rho}$, $f_{\eta_c}<f_{J/\psi}$.  We therefore choose to constrain $m_{b}$, $\Lambda_{\rm uv}^{\eta_{b}}$ via known experimental results \cite{Tanabashi:2018oca}:
\begin{equation}
\label{fetab}
f_{\eta_b} = f_{\Upsilon} [ f_{\eta_c}/f_{J/\psi} ] = 0.41(2)\,.
\end{equation}

It is worth remarking here that our fitted values of $m_c$, $m_b-m_c$ are commensurate with QCD estimates and the computed results for $M_{c,b}$ are aligned with typical values of the heavy-quark pole masses \cite{Tanabashi:2018oca}.

Another useful feature of Eq.\,\eqref{alphaLambda} is that it enables us to implement an important physical constraint, \emph{viz}.\ any increase in the momentum-space extent of a hadron wave function must be accompanied by a commensurate decrease in the effective coupling between the constituents so as to avoid critical over-binding.  Our analysis yields
\begin{equation}
\Lambda_{\rm uv}(s=m_{0^-}^2) \stackrel{m_{0^-} \geq m_K}{=} 0.83 \ln[2.79 + s/(4.66 \Lambda_{\rm ir})^2 ]\,;
\end{equation}
and via Eq.\,\eqref{alphaLambda}, an evolution of the quark-antiquark coupling that is well approximated by
\begin{align}
\label{alphaIRMass}
\alpha_{\rm IR}(s) \stackrel{m_{0^-} \geq m_K}= \frac{0.047\,\alpha_{\rm IR}(m_K^2) }{\ln[1.04 + s/(21.77 \Lambda_{\rm ir})^2]}\,.
\end{align}
The origin of these outcomes is plain: the decay-constant integral diverges logarithmically with increasing $\Lambda_{\rm uv}$; and $\alpha_{\rm IR}$ flows to compensate for analogous behaviour in the Bethe-Salpeter kernel and thereby maintain the given meson's mass.


\section{Meson Spectrum}
\label{SecMeson}
It is now possible to compute the masses, Bethe-Salpeter amplitudes and leptonic decay constants of a wide array of ground-state pseudoscalar and vector mesons with valence-quark content $(f\bar g)$.

The BSE for such pseudoscalar mesons is given in Eq.\,\eqref{bsefinalE}, with the associated leptonic decay constant computed from Eq.\,\eqref{ffg}.
Turning to $(f\bar g)$ vector mesons, the most general form of the Bethe-Salpeter amplitude supported by a  RL analysis of the contact interaction is
\begin{equation}
\Gamma_\mu^{1^-}(Q) = \gamma_\mu^{\perp} E_{1^-}(Q)\,,
\end{equation}
where $Q\cdot \gamma_\mu^{\perp} = 0$.  This dimensionless constant, $E_{1^-}(Q)$, is determined by solving the BSE obtained via straightforward generalisation of Eqs.\,(18)\,-\,(20) in Ref.\,\cite{Chen:2012qr}.  The associated leptonic decay constant is computed from
\begin{align}
\nonumber
f_{1^-}\,m_{1^-} & = \frac{3}{4\pi^2} \int_0^1d\alpha\,
[M_f M_{g} - \alpha\hat\alpha Q^2 - \omega_{fg}(\alpha,Q^2)]\\
& \qquad \times
\overline{\mathpzc C}_1^{\rm iu}(\omega_{f g}(\alpha,Q^2))
E_{1^-}(Q)\,,
\end{align}
where $\overline{\mathpzc C}_1^{\rm iu}$ is defined in Appendix~\ref{AppFormulae}.

\begin{table}[t]
\caption{\label{Tab:MesonSpectrum}
Computed Bethe-Salpeter amplitudes, masses and decay constants of pseudoscalar and vector mesons.  The underlined entries, repeated from Table~\ref{Tab:DressedQuarks}, were used to fit the interaction strength, current-quark masses and ultraviolet cutoffs.
Empirical masses, where known, are taken from Ref.\,\cite{Tanabashi:2018oca} ; $m_{B^\ast_c}$ is from Ref.\,\cite{Mathur:2018epb}; and for those decay constants not known experimentally, we typically quote lQCD results \cite{Becirevic:1998ua, Chiu:2007bc, Davies:2010ip, McNeile:2012qf, Donald:2012ga, Colquhoun:2015oha}.
(Dimensioned quantities in GeV.)}
\begin{center}
\begin{tabular*}
{\hsize}
{
c@{\extracolsep{0ptplus1fil}}|
c@{\extracolsep{0ptplus1fil}}
l@{\extracolsep{0ptplus1fil}}
c@{\extracolsep{0ptplus1fil}}
c@{\extracolsep{0ptplus1fil}}
|c@{\extracolsep{0ptplus1fil}}
l@{\extracolsep{0ptplus1fil}}}\hline\hline
meson, $M\ $ & $E_M\ $ & $\ F_M\ $ & $m_M^{\rm CI}\ $ &   $f_M^{\rm CI}\ $ &  $m_M^{\rm e/l}\ $ & $\ f_{M}^{\rm e/l}\ $ \\\hline
$\pi\ $ & $3.59\ $ & $\ 0.47\ $ & $\underline{0.14}\ $ & $\underline{0.10}\ $ & $0.14\ $ & $\ 0.092\ $\\
$K\ $ & $3.82\ $ & $\ 0.56\ $ & $\underline{0.50}\ $ & $\underline{0.11}\ $ & $0.50\ $ & $\ 0.11 $\\
$\rho\ $ & $1.53\ $ &  & $0.93\ $ & $0.13\ $ & $0.78\ $ & $\ 0.15\ $\\
$K^\ast$ & $1.63\ $ &  & $1.03\ $ & $0.12\ $ & $0.89\ $ & $\ 0.16\ $\\
$\phi\ $ & $1.74\ $ &  & $1.13\ $ & $0.12\ $ & $1.02\ $ & $\ 0.17\ $\\\hline
$D\ $ & $3.11\ $& $\ 0.36\ $& $1.92\ $& $0.16\ $ & $1.87\ $ & $\ 0.15(1)\ $ \\ 
$D_s\ $ & $3.25\ $ & $\ 0.49\ $ & $2.01\ $ & $0.17\ $ & $1.97\ $ & $\ 0.18\ $ \\ 
$D^\ast\ $ & $1.21\ $ & & $2.14\ $ &$0.15\ $ & $2.01\ $ & $\ 0.17(1)\ $ \\ 
$D^\ast_s\ $ & $1.23\ $& & $2.23\ $ & $0.16\ $& $2.11\ $ & $\ 0.19\ $ \\ 
$\eta_c\ $ & $3.28\ $ & $\ 0.73\ $ & $\underline{2.98}\ $ & $\underline{0.24}\ $ & $2.98\ $ & $\ 0.24\ $ \\ 
$J/\psi\ $ & $1.21\ $ & & $3.19\ $ & $0.20\ $ & $3.10\ $ & $\ 0.29\ $ \\ 
$B\ $ & $1.67\ $ & $\ 0.095\ $ & $5.41\ $ &$0.17\ $  & $5.30\ $ & $\ 0.14(2)\ $ \\ 
$B^\ast\ $ & $0.70\ $ & & $5.46\ $ & $0.16\ $ & $5.33\ $ & $\ 0.12\ $ \\ 
$B_s\ $ & $1.79\ $ & $\ 0.14\ $ & $5.50\ $ & $0.18\ $ & $5.37\ $ & $\ 0.16\ $ \\ 
$B^\ast_s\ $ & $0.71\ $ & & $5.56\ $ & $0.16\ $ & $5.42\ $ & $\ 0.15(1)\ $ \\ 
$B_c\ $ & $3.38\ $ & $\ 0.61\ $ & $6.28\ $ & $0.27\ $ & $6.28\ $ & $\ 0.35\ $ \\ 
$B^\ast_c\ $ & $1.37\ $ & & $6.38\ $ & $0.23\ $ & $6.33\ $ & $\ 0.30(1)\ $ \\ 
$\eta_b\ $ & $3.18\ $ & $\ 0.81\ $ & $\underline{9.40}\ $ & $\underline{0.41}\ $ & $9.40\ $ & $\ 0.41(2)\ $ \\ 
$\Upsilon\ $ & $1.50\ $ & & $9.49\ $ & $0.38\ $ & $9.46\ $ & $\ 0.46\ $  
\\\hline\hline
\end{tabular*}
\end{center}
\end{table}

Our computed results are gathered in Table~\ref{Tab:MesonSpectrum}.  For those systems with mass $m_{0^-}\geq m_K$ we fix $\Lambda_{\rm uv}(m_{0^-}^2)$, $\alpha_{\rm IR}(m_{0^-}^2)$ using Eqs.\,\eqref{alphaLambda}, \eqref{alphaIRMass}; hence, every contact-interaction result in the table is a prediction except those in the four rows with underlined entries.  Notably, the $F_{0^-}$ (pseudovector) component of each pseudoscalar meson is nonzero: on average, it is 15(6)\% of the $E_{0^-}$ (pseudoscalar) piece.  Hence, the pseudovector component is quantitatively important in all cases.

\begin{figure}[t]
\includegraphics[clip, width=0.45\textwidth]{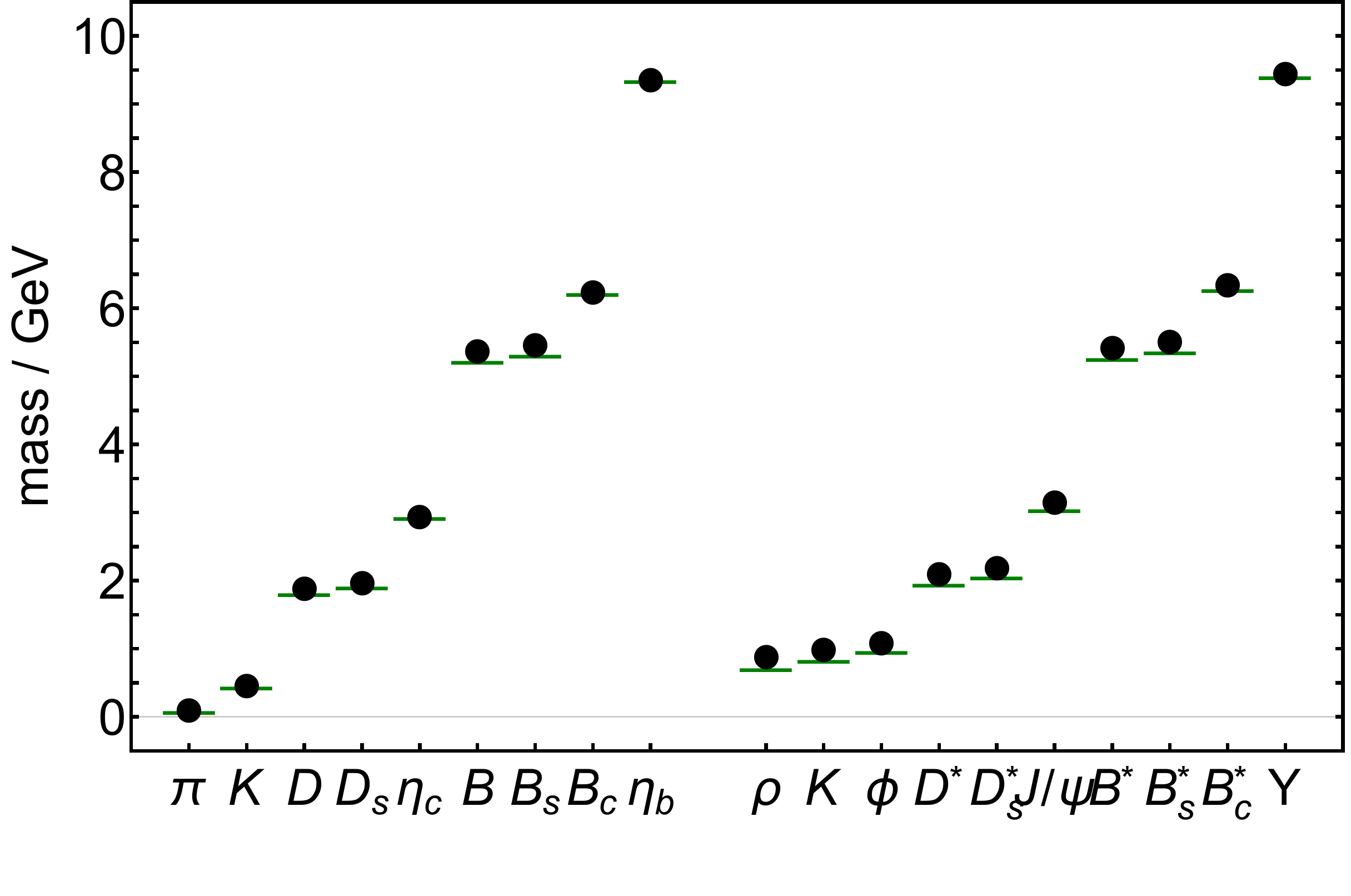}
\includegraphics[clip, width=0.45\textwidth]{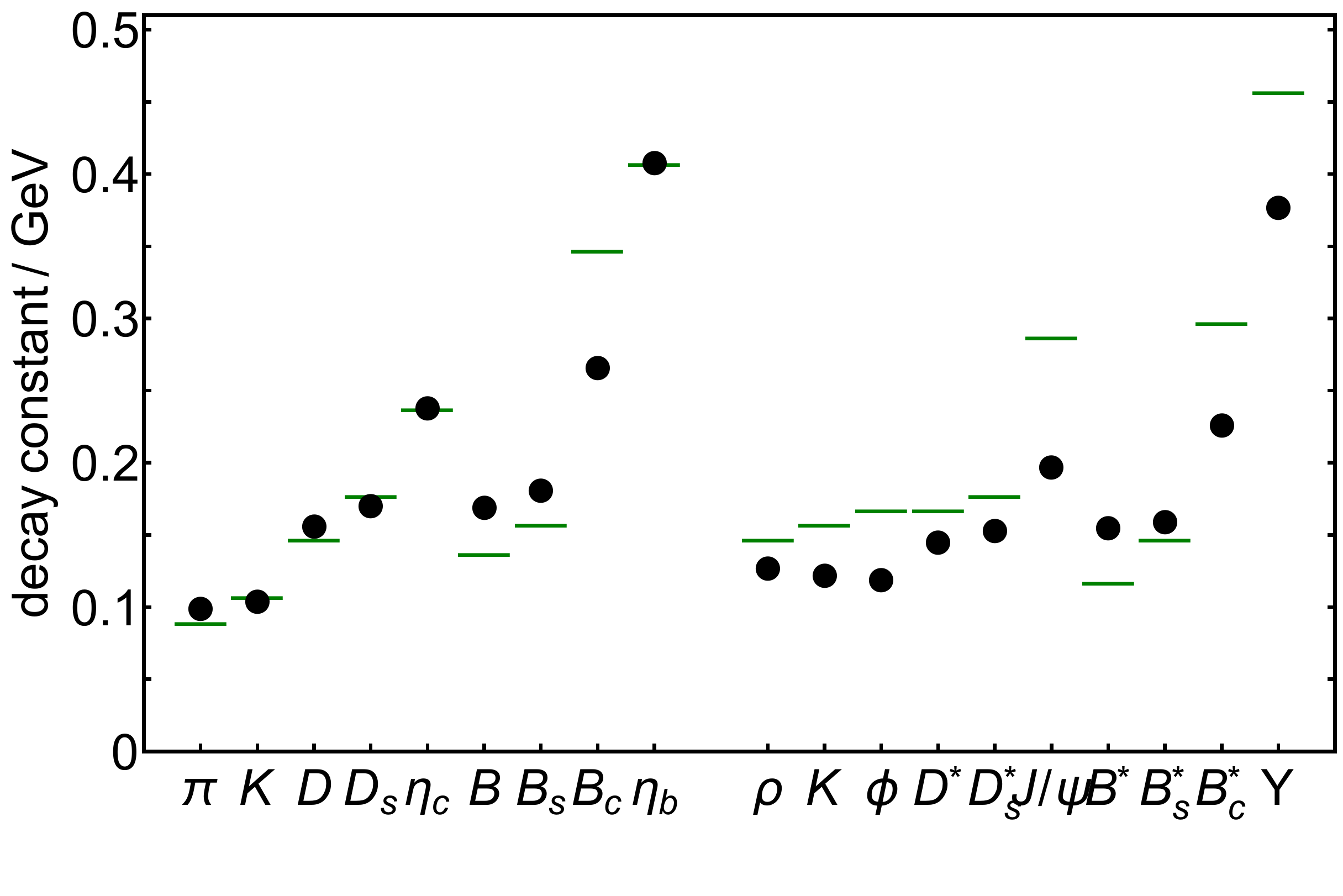}
\caption{\label{FigMeson}
\emph{Upper panel}.  Comparison between contact-in\-ter\-ac\-tion predictions for meson masses and experiment \cite{Tanabashi:2018oca}. ($m_{B^\ast_c}$ is from Ref.\,\cite{Mathur:2018epb}.)
\emph{Lower panel}.  Contact-interaction predictions for meson leptonic decay constants and experiment \cite{Tanabashi:2018oca}, where known, and lQCD otherwise \cite{Becirevic:1998ua, Chiu:2007bc, Davies:2010ip, McNeile:2012qf, Donald:2012ga, Colquhoun:2015oha}.
In both panels, contact-interaction predictions are depicted as (black) circles and comparison values by (green) bars.
}
\end{figure}

To aid with understanding the comparisons in Table~\ref{Tab:MesonSpectrum}, we also depict them in Fig.\,\ref{FigMeson}.
Considering the upper panel, it is evident that our treatment of the contact interaction delivers good estimates for the masses of regular ground-state flavour-$SU(5)$ mesons: aspects of its symmetry-preserving formulation lead to an overestimate in each case \cite{Roberts:2010rn}, but the mean relative-difference between theory and experiment/lQCD is just 5(5)\%.  Furthermore, the computed masses fit neatly within a pattern prescribed by equal spacing rules (ESRs) \cite{Okubo:1961jc, GellMann:1962xb, Qin:2018dqp, Chen:2019fzn, Qin:2019hgk}, \emph{e.g}.\
{\allowdisplaybreaks
\begin{subequations}
\label{ESRexamples}
\begin{align}
m_{K^\ast} & = [ m_\phi + m_\rho ]/2 \,,\\
m_{B^\ast_s}-m_{B^\ast} & = m_{B_s}-m_{B}  \\
 & = m_{D^\ast_s}-m_{D^\ast}  \\
 & = m_{D_s}-m_{D}\,,\\
m_{B^\ast_c} - m_{B^\ast_s} & \approx m_{B_c} - m_{B_s} \,, \\
[m_{\Upsilon}-m_{J/\psi}]/2 & \approx [m_{\eta_b}-m_{\eta_c}]/2 \\
 & \approx m_{B^\ast_s} - m_{D^\ast_s} = m_{B_s} - m_{D_s} \,.
\end{align}
\end{subequations}}

The comparison between our contact-interaction predictions for meson leptonic decay constants and experiment/lQCD is depicted in the lower panel of Fig.\,\ref{FigMeson}.  Physically, a meson's leptonic decay constant is sensitive to ultraviolet physics (the constituents annihilate at a single point in spacetime).  This is expressed in the QCD expression for the decay constant through the appearance of a logarithmic ultraviolet divergence, which is controlled by the dressed-quark wave-function renormalisation constant \cite{Maris:1997hd, Chang:2008ec}.  Given these features, it is not surprising that the cutoff-regularised contact-interaction provides a poorer description of decay constants than it does of masses.  Nevertheless, the picture is tolerable: general trends are reproduced and the mean-absolute-relative-difference between the entries in columns 4 and 6 of Table~\ref{Tab:MesonSpectrum} is 18(9)\%.\footnote{If Eqs.\,\eqref{alphaLambda}, \eqref{alphaIRMass} are not implemented, the description of the decay constants is bad.  A hint for this is found in a comparison between the CI-calculated trend for $f_{\rho}$, $f_{K^\ast}$, $f_{\phi}$ and experiment.}    Owing to peculiarities of the contact-interaction's symmetry-preserving formulation \cite{Roberts:2010rn}, the description is better for pseudoscalar mesons than for vector mesons.  Notwithstanding these observations, analogues of Eqs.\,\eqref{ESRexamples} are applicable, \emph{e.g}.\
{\allowdisplaybreaks
\begin{subequations}
\label{ESRfexamples}
\begin{align}
f_{K^\ast} & \approx [ f_\phi + f_\rho ]/2 \,,\\
f_{B^\ast_s}-f_{B^\ast}  & \approx f_{D^\ast_s}-f_{D^\ast}  \\
f_{B_s}-f_{B}  & \approx f_{D_s}-f_{D}\,,\\
f_{B^\ast_c} - f_{B^\ast_s} & \approx f_{B_c} - f_{B_s} \,, \\
[f_{\Upsilon}-f_{J/\psi}]/2 & \approx [f_{\eta_b}-f_{\eta_c}]/2 \\
 & \approx  f_{B_s} - f_{D_s} \,.
\end{align}
\end{subequations}}

\section{Spectrum of Diquark Correlations}
\label{SecDiquarks}
In solving for the spectrum of flavour-$SU(5)$ ground-state baryons, we use a quark-diquark approximation to the Faddeev equation.\footnote{It is worth reiterating that the diquarks described herein are fully dynamical and appear in a Faddeev kernel which requires their continual breakup and reformation.  Consequently, they are vastly different from the static, pointlike diquarks introduced originally \cite{Anselmino:1992vg} in an attempt to solve the so-called ``missing resonance'' problem \cite{Aznauryan:2011ub}.}
It is therefore necessary to know the masses and amplitudes for all diquark correlations that can exist in these systems.
Fortunately, having solved for the spectrum of flavour-$SU(5)$ mesons, it is straightforward to compute these diquark quantities because the RL BSE for a $J^P$ diquark is obtained directly from that for a $J^{-P}$ meson simply by multiplying the meson kernel by a factor of $1/2$ \cite{Cahill:1987qr}.  (The flipping of the sign in parity occurs because parity is opposite for fermions and antifermions.)  For instance, the Bethe-Salpeter equation for a scalar $[fg]$ diquark is
\begin{align}
\nonumber
\Gamma^C_{[fg]}&(Q) =  - \frac{1}{2} \frac{16 \pi}{3} \frac{\alpha_{\rm IR}}{m_G^2}\\
& \times
\int \! \frac{d^4t}{(2\pi)^4} \gamma_\mu S_f(t+Q) \Gamma^C_{[fg]}(Q) S_g(t) \gamma_\mu \,,
\label{LBSEqq}
\end{align}
where the correlation amplitude is $\Gamma_{[fg]}(Q)$ and
\begin{subequations}
\begin{align}
\Gamma^C_{[fg]}(Q) & := \Gamma_{[fg]}(Q)C^\dagger\\
& = \gamma_5 \left[ i E_{[fg]}(Q) + \frac{1}{2 M_R} \gamma\cdot Q F_{[fg]}(Q)\right]\,,
\end{align}
\end{subequations}
with $C=\gamma_2\gamma_4$ being the charge-conjugation matrix.  The canonical normalisation conditions are similarly amended, with the multiplicative factor being $2/3$ in this case (see, \emph{e.g}.\ Eqs.\,(24), (25) in Ref.\,\cite{Chen:2012qr}).

\begin{table}[t]
\caption{\label{diquarkspectrum}
Computed masses and amplitudes for the diquark correlations:  $[fg]$ indicates a $J^P=0^+$ diquark, antisymmetric under $f\leftrightarrow g$; and $\{fg\}$ indicates a $J^P=1^+$ diquark, symmetric under $f\leftrightarrow g$.
(Masses listed in GeV.)}
\begin{center}
\begin{tabular*}
{\hsize}
{
c@{\extracolsep{0ptplus1fil}}|
l@{\extracolsep{0ptplus1fil}}
|l@{\extracolsep{0ptplus1fil}}
l@{\extracolsep{0ptplus1fil}}}\hline\hline
diquark$\ $ & mass$\ $ & $\ E\ $ & $\ F\ $ \\ \hline
$[ud]$ & $0.77\ $ & $2.74\ $ & $0.31\ $ \\
$[us]$ & $0.93\ $ & $2.88\ $ & $0.39\ $ \\
$[uc]$ & $2.15\ $ & $1.97\ $ & $0.22\ $ \\
$[sc]$ & $2.26\ $ & $1.99\ $ & $0.29\ $ \\
$[ub]$ & $5.51\ $ & $1.05\ $ & $0.059\ $ \\
$[sb]$ & $5.60\ $ & $1.05\ $ & $0.083\ $ \\
$[cb]$ & $6.48\ $ & $1.42\ $ & $0.25\ $ \\
$\{uu\}$ & $1.06\ $ & $1.31\ $ & \\
$\{us\}$ & $1.16\ $ & $1.36\ $ & \\
$\{ss\}$ & $1.26\ $ & $1.43\ $ & \\
$\{uc\}$ & $2.24\ $ & $0.89\ $ & \\
$\{sc\}$ & $2.34\ $ & $0.87\ $ & \\
$\{cc\}$ & $3.30\ $ & $0.69\ $ & \\
$\{ub\}$ & $5.53\ $ & $0.51\ $ & \\
$\{sb\}$ & $5.62\ $ & $0.50\ $ & \\
$\{cb\}$ & $6.50\ $ & $0.62\ $ & \\
$\{bb\}$ & $9.68\ $ & $0.48\ $ & \\
\hline\hline
\end{tabular*}
\end{center}
\end{table}

Following the approach indicated above and using the parameters determined in Sec.\,\ref{SecMeson}, one obtains the diquark masses and amplitudes listed in Table~\ref{diquarkspectrum}.\footnote{In the Bethe-Salpeter equation for a given $J^P$ diquark correlation we employ the values of $\alpha_{\rm IR}$, $\Lambda_{\rm uv}$ associated with its $J^{-P}$ meson partner, a scheme consistent with RL studies using realistic interactions \cite{Maris:2002yu, Maris:2004bp}.}
Evidently, the antisymmetric combination of any two quark flavours is always lighter than the symmetric combination and the pattern of masses can be understood in terms of equal spacing rules, just as was the case for mesons.

It is pertinent to remark here that RL truncation generates asymptotic diquark states.  Such states are not observed and their appearance is an artefact of the truncation.  Higher-order terms in the quark-quark scattering kernel, whose analogue in the quark-antiquark channel do not materially affect the properties of vector and flavour non-singlet pseudoscalar mesons, ensure that QCD's quark-quark scattering matrix does not exhibit singularities which correspond to asymptotic diquark states \cite{Bender:2002as, Bhagwat:2004hn}.   Studies with kernels that exclude diquark bound states nevertheless support a physical interpretation of the masses, $m_{(fg)^{\!J^P}}$, obtained using the rainbow-ladder truncation, \emph{viz}.\ the quantity $\ell_{(fg)^{\!J^P}}:=1/m_{(fg)^{\!J^P}}$ may be interpreted as a range over which the diquark correlation can propagate before fragmentation.

\begin{figure}[t]
\includegraphics[clip, width=0.45\textwidth]{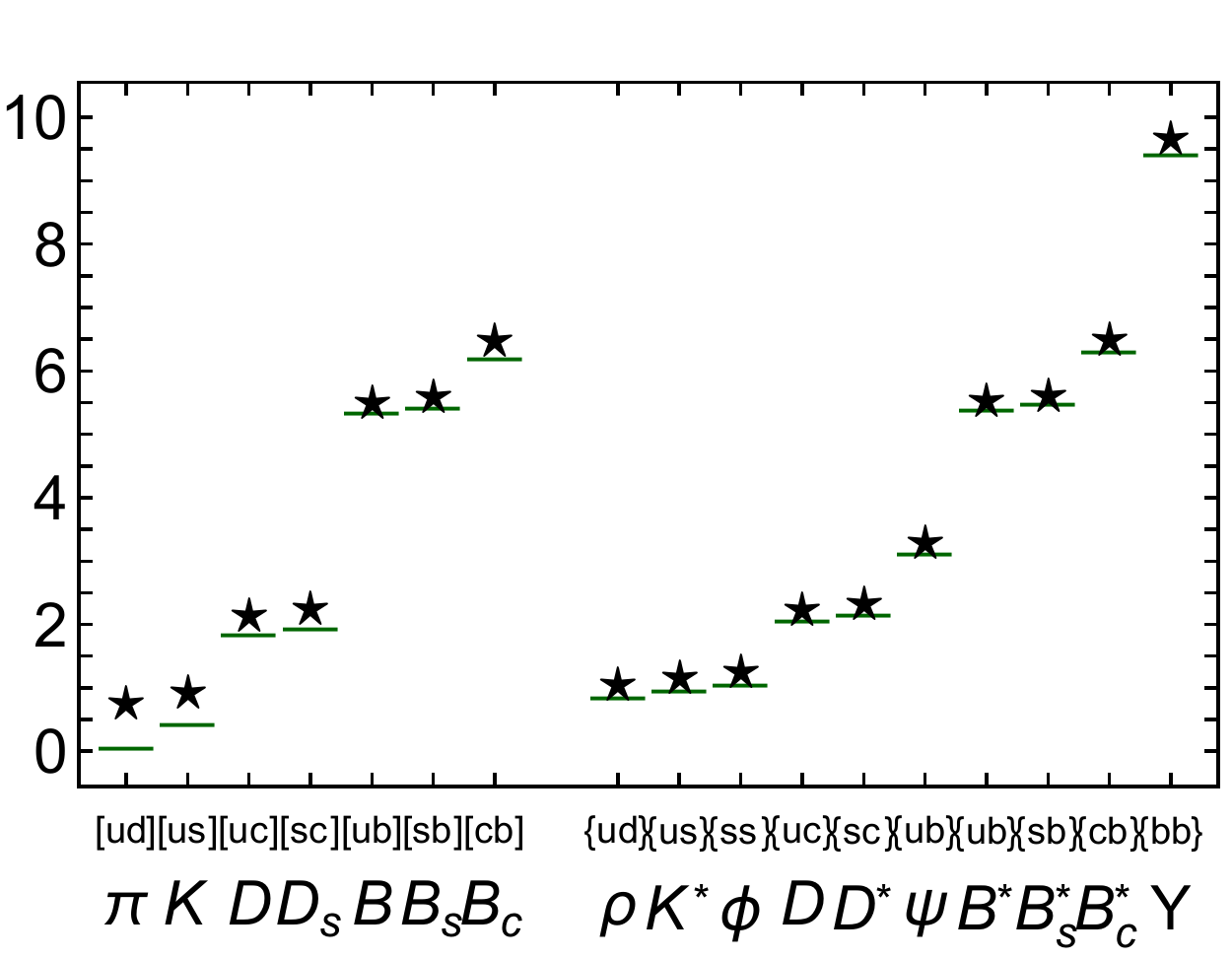}
\caption{\label{Figqq}
Comparison between computed masses of diquark correlations and their symmetry-related meson counterparts: diquarks -- (black) stars and mesons -- (green) bars.
}
\end{figure}

In Fig.\,\ref{Figqq} we compare the diquark masses with those of their partner mesons: the level ordering of diquark correlations is precisely the same as that for mesons and the meson mass bounds the partner diquark's mass from below.  Moreover, in all cases, except the $l,s$ scalar diquarks, the mass of the diquark's partner meson is a fair guide to the diquark's mass: the mean difference is $0.13(6)\,$GeV.

The light-quark scalar diquark channels are atypical owing to dynamical chiral symmetry breaking and the Nambu-Goldstone boson character of the partner pseudoscalar mesons.  Notably, in a two-color version of QCD, the scalar diquark is also a Nambu-Goldstone mode \cite{Roberts:1996jx, Bloch:1999vk}, a long-known result of Pauli-G\"ursey symmetry \cite{Pauli:1957, Gursey:1958}.  A memory of the symmetry persists in the three-color theory and is evident here in low masses for the $l,s$ scalar diquarks, even though they are nevertheless split widely from the true Nambu-Goldstone mesons.

In constructing baryon Faddeev equations, the canonically normalised diquark Bethe-Salpeter amplitudes are crucial because they determine the strength of the correlation within a given baryon.  We list them in Table~\ref{diquarkspectrum}.  Notably, the leading amplitudes associated with flavour-antisymmetric correlations are a factor of $2.2(1)$ larger than their flavour-symmetric counterparts.  This serves to amplify the preference for the lighter flavour-antisymmetric correlations in a given $J^P=1/2^+$ baryon because it is the amplitude-squared which appears in the Faddeev equations.  As we shall see, this preference can be overcome in $1/2^+$ baryons whose valence-quarks possess very different masses.  (The $J^P=3/2^+$ baryons considered herein possess flavour-exchange symmetries which prevent the presence of $0^+$ $[fg]$ correlations.)

\section{Baryon Spectrum}
\label{SecBaryons}
\subsection{Structure of Faddeev Amplitudes}
\begin{figure}[t]
\centerline{%
\includegraphics[clip, width=0.45\textwidth]{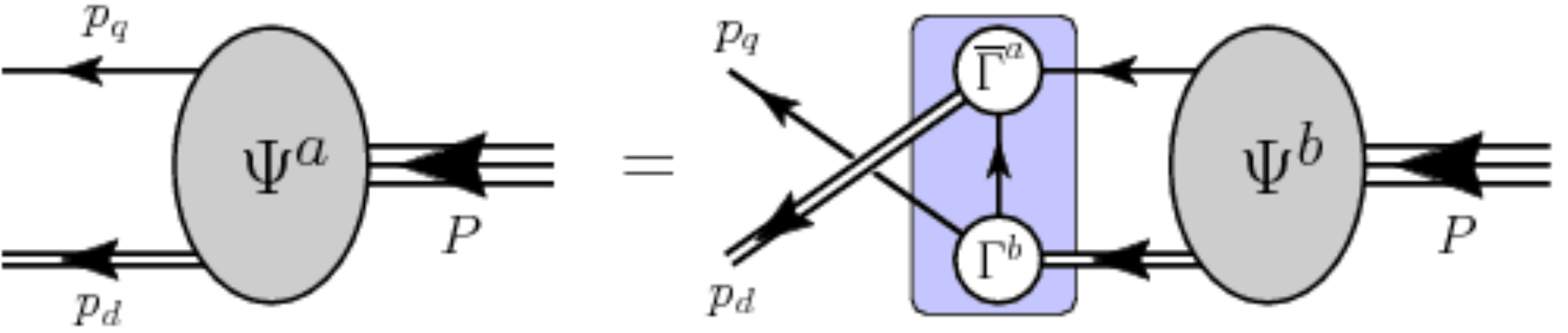}}
\caption{\label{figFaddeev}
Poincar\'e covariant Faddeev equation: a linear integral equation for the matrix-valued function $\Psi$, being the Faddeev amplitude for a baryon of total momentum $P= p_d+p_q$, which expresses the relative momentum correlation between the dressed-quarks and -nonpointlike-diquarks within the baryon.  The shaded rectangle demarcates the kernel of the Faddeev equation:
\emph{single line}, dressed-quark propagator (Sec.\,\ref{SecIS}); $\Gamma$,  diquark correlation amplitude (Sec.\,\ref{SecDiquarks}); and \emph{double line}, diquark propagator (Eqs.\,\eqref{scalarqqprop}, \eqref{qqavprop}). }
\end{figure}

We use the Faddeev equation depicted in Fig.\,\ref{figFaddeev} to compute the spectrum of ground-state flavour-$SU(5)$ $J^P=1/2^+, 3/2^+$ baryons, following the patterns of analysis described in Ref.\,\cite{Chen:2012qr}.  Note, then, that a spin-$1/2$ baryon may be represented by a Faddeev amplitude \cite{Cahill:1988dx}
\begin{equation}
\label{PsiBaryon}
\Psi = \Psi_1 + \Psi_2 + \Psi_3  \,,
\end{equation}
where the subscript identifies the bystander quark and, \emph{e.g}.\ $\Psi_{1,2}$ are obtained from $\Psi_3$ by a cyclic permutation of all the quark labels.  We employ the simplest realistic representation of $\Psi$, so that any member of the flavour-$SU(5)$ multiplet which generalises the $SU(3)$-octet of baryons is composed from a sum of scalar and axial-vector diquark correlations:
\begin{equation}
\label{Psi} \Psi_3(p_j,\alpha_j,\varphi_j) = {\cal N}_{\;\Psi_3}^{0^+} + {\cal N}_{\;\Psi_3}^{1^+},
\end{equation}
with $(p_j,\alpha_j,\varphi_j)$ the momentum, spin and flavour labels of the quarks constituting the bound state, and $P=(p_1+p_2)+p_3=p_d+p_q$ the system's total momentum.  (\emph{N.B}.\ Negative-parity diquark correlations play no material role in positive-parity ground-state baryons \cite{Eichmann:2016hgl, Lu:2017cln, Chen:2017pse, Chen:2019fzn}.)

Using the flavour-matrices defined in Eqs.\,\eqref{flavourarrays1}, the scalar diquark piece in Eq.\,(\ref{Psi}) can be written
\begin{align}
\nonumber
& {\cal N}_{\;\Psi_3}^{0^+}(p_j,\alpha_j,\varphi_j)  = \\
& \sum_{[\varphi_1\varphi_2]\varphi_3\in \Psi} \bigg[{\tt t}^{[\varphi_1\varphi_2]}\,
\Gamma_{[\varphi_1\varphi_2]}^{0^+}(\frac{1}{2}p_{[12]};K)\bigg]_{\alpha_1
\alpha_2}^{\varphi_1 \varphi_2}\, \nonumber \\
& \quad \times \Delta_{[\varphi_1\varphi_2]}^{0^+}(K) \,[{\cal S}^{\Psi}(\ell;P) u^\Psi(P)]_{\alpha_3}^{\varphi_3} ,
\label{calS}
\end{align}
where:
$K= p_1+p_2=: p_{\{12\}}$,
$p_{[12]}= p_1 - p_2$, $\ell := (-p_{\{12\}} + 2 p_3)/3$;
\begin{equation}
\label{scalarqqprop}
\Delta_{[\varphi_1\varphi_2]}^{0^+}(K) = \frac{1}{K^2+m_{[\varphi_1\varphi_2]}^2}
\end{equation}
is a propagator for the scalar diquark formed from quarks $1$ and $2$, with $m_{[\varphi_1\varphi_2]}$ being the associated mass-scale and $\Gamma_{[\varphi_1\varphi_2]}^{0^+}\!$ the canonically-normalised Bethe-Salpeter amplitude describing the correlation strength between the quarks, all computed in Sec.\,\ref{SecDiquarks}; ${\mathpzc S}$, a $4\times 4$ Dirac matrix, describes the relative quark-diquark momentum correlation within the baryon; and the spinor satisfies
\begin{equation}
(i\gamma\cdot P + M_\Psi)\, u^\Psi(P) =0= \bar u^\Psi(P)\, (i\gamma\cdot P + M_\Psi)\,,
\end{equation}
with $M_\Psi$ the baryon mass obtained by solving the Faddeev equation.  The flavour structure of $1/2^+$ baryons is expressed in Eqs.\,\eqref{uColumn}.

The axial-vector part of Eq.\,(\ref{Psi}) is
\begin{align}
& {\cal N}_{\;\Psi_3}^{1^+}(p_j,\alpha_j,\varphi_j) = \nonumber \\
& \sum_{\{\varphi_1\varphi_2\}\varphi_3\in \Psi} \bigg[{\tt t}^{\{\varphi_1\varphi_2\}}\,
\Gamma_{\mu\{\varphi_1\varphi_2\}}^{1^+}(\frac{1}{2}p_{[12]};K)\bigg]_{\alpha_1
\alpha_2}^{\varphi_1 \varphi_2}\, \nonumber \\
& \quad \times \Delta_{1^+\mu\nu}^{\{\varphi_1\varphi_2\}}(K) \,[{\cal A}_\nu^{\Psi}(\ell;P) u^\Psi(P)]_{\alpha_3}^{\varphi_3} ,
\label{calA}
\end{align}
where
\begin{equation}
\Delta_{1^+ \mu\nu}^{\{\varphi_1\varphi_2\}}(K) = \frac{1}{K^2+m_{m^2_{\{\varphi_1 \varphi_2\}}}} \, \left(\delta_{\mu\nu} + \frac{K_\mu K_\nu}{m^2_{\{\varphi_1 \varphi_2\}}}\right) ,
\label{qqavprop}
\end{equation}
is a propagator for the axial-vector diquark formed from quarks $1$ and $2$ and the other elements in Eq.\,\eqref{calA} are obvious analogues of those in Eq.\,\eqref{calS}.

Regarding baryons in the flavour-$SU(5)$ generalisation of the usual decuplet, one may write
\begin{equation}
\label{DecupletFA}
\Psi_3^{\wedge}(p_i,\alpha_i,\varphi_i) = {\cal D}_{\Psi_3^{\wedge}}^{1^+}(p_j,\alpha_j,\varphi_j),
\end{equation}
with
\begin{align}
& {\cal D}_{\,\Psi_3^{\wedge}}^{1^+}(p_j,\alpha_j,\varphi_j)= \nonumber \\
& \! \sum_{\{\varphi_1\varphi_2\}\varphi_3\in \Psi^{\wedge}}
\bigg[{\tt t}^{\{\varphi_1\varphi_2\}}\,
\Gamma_{\mu\{\varphi_1\varphi_2\}}^{1^+}(\frac{1}{2}p_{[12]};K)\bigg]_{\alpha_1
\alpha_2}^{\varphi_1 \varphi_2} \rule{-1ex}{0ex} \nonumber \\
& \quad \times
\Delta_{1^+\mu\nu}^{\{\varphi_1\varphi_2\}}(K) \,[{\cal D}_{\nu\rho}^{\Psi^{\wedge}}(\ell;P) u_\rho^{\Psi^{\wedge}}(P)]_{\alpha_3}^{\varphi_3} ,
\label{calD}
\end{align}
where $u_\rho^{\Psi^{\wedge}}(P)$ is a Rarita-Schwinger spinor and, as with $1/2^+$ baryons, in constructing the Faddeev equations we focus on that member of each isospin multiplet which has maximum electric charge, \emph{viz}.\ the $20$ states in Eqs.\,\eqref{uDeltaColumn}.

\begin{table*}[t]
\caption{\label{ResultsHalf}
Computed mass and Faddeev amplitude for each ground-state flavour-$SU(5)$ $J^P=1/2^+$ baryon: the last column highlights the baryon's dominant spin-flavour correlation.  Empirical mass values are taken from Ref.\,\cite{Tanabashi:2018oca}; where they are absent, lQCD results are listed \cite{Brown:2014ena, Mathur:2018epb} and indicated by ``$\ast$''.  The parenthesised label beside a baryon's name indicates the equation which specifies the associated Faddeev amplitude's spin-flavour vector.
The subscript on the axial-vector entry specifies the Dirac structure in Eq.\,\eqref{calAcontact}.
(Mass in GeV.)
}
\begin{center}
\begin{tabular*}
{\hsize}
{
l@{\extracolsep{0ptplus1fil}}
|l@{\extracolsep{0ptplus1fil}}
l@{\extracolsep{0ptplus1fil}}
|l@{\extracolsep{0ptplus1fil}}
l@{\extracolsep{0ptplus1fil}}
l@{\extracolsep{0ptplus1fil}}
l@{\extracolsep{0ptplus1fil}}
l@{\extracolsep{0ptplus1fil}}
l@{\extracolsep{0ptplus1fil}}
|c@{\extracolsep{0ptplus1fil}}}\hline
Baryon & $M^{e/l}\ $ & $M^{\rm CI}$ & $\ s^{{\rm r}_1}$ & $\ s^{{\rm r}_2}\ $ & $\ a_1^{{\rm r}_2}\ $ & $\ a_2^{{\rm r}_2}\ $  & $\ a_1^{{\rm r}_3}\ $ & $\ a_2^{{\rm r}_3}\ $
& dom.\,corr.$\ $ \\\hline
$p\ $ \eqref{B5a}\,
& $0.94\ $ & $0.94\ $ & $0.89\ $ & & $-0.35\ $ & $-0.14\ $ & $\phantom{-}0.25\ $ & $\phantom{-}0.098\ $ & $[ud]u\ $ \\
$\Lambda\ $ \eqref{B5d}\,
& $1.12\ $ & $1.06\ $ & $0.67\ $ & $\phantom{-}0.59\ $&  &  & $-0.42\ $ & $-0.16\ $ &
$[ud]s\ $\\
$\Sigma\ $ \eqref{B5b}\,
& $1.19\ $ & $1.20\ $ & $0.87\ $ & & $-0.42\ $ & $\phantom{-}0.004\ $ & $\phantom{-}0.25\ $ & $\phantom{-}0.071\ $ & $[us]u\ $\\
$\Xi\ $ \eqref{B5c}\,
& $1.32\ $ & $1.24\ $ & $0.90\ $ & & $-0.29\ $ & $-0.028\ $ & $\phantom{-}0.31\ $ & $\phantom{-}0.11\ $ & $[us]s\ $\\
$\Lambda_c$ \eqref{B5g}\,
& $2.29\ $ & $2.50\ $ & $0.21\ $ & $\phantom{-}0.86\ $&  &  & $-0.35\ $ & $-0.32\ $ &
$[uc]d-[dc]u\ $\\
$\Sigma_c $ \eqref{B5e}\,
& $2.45\ $ & $2.53\ $ & $0.48\ $ & & $-0.21\ $ & $\phantom{-}0.84\ $ & $\phantom{-}0.090\ $ & $\phantom{-}0.064\ $ & $\{uu\}c\ $\\
$\Xi_c$ \eqref{Xic}\,
& $2.47\ $ & $2.66\ $ & $0.22\ $ & $\phantom{-}0.84\ $&  &  & $-0.36\ $ & $-0.34\ $ &
$[uc]s-[sc]u\ $\\
$\Xi_c^\prime $ \eqref{Xicprime}\,
& $2.58\ $ & $2.68\ $ & $0.50\ $ & & $-0.22\ $ & $\phantom{-}0.83\ $ & $\phantom{-}0.093\ $ & $\phantom{-}0.061\ $ & $\{us\}c\ $ \\
$\Omega_c $ \eqref{B5f}\,
& $2.70\ $ & $2.83\ $ & $0.51\ $ & & $-0.22\ $ & $\phantom{-}0.82\ $ & $\phantom{-}0.097\ $ & $\phantom{-}0.058\ $ & $\{ss\}c\ $\\
$\Lambda_b$ \eqref{B5l}\,
& $5.62\ $ & $5.74\ $ & $0.13\ $ & $\phantom{-}0.93\ $&  &  & $-0.31\ $ & $-0.13\ $&
$[ub]d-[db]u\ $\\
$\Sigma_b $ \eqref{B5j}\,
& $5.81\ $ & $5.85\ $ & $0.30\ $ & & $-0.12\ $ & $\phantom{-}0.94\ $ & $\phantom{-}0.041\ $ & $\phantom{-}0.087\ $& $\{uu\}b\ $\\
$\Xi_b$ \eqref{B5m}\,
& $5.79\ $ & $5.88\ $ & $0.13\ $ & $\phantom{-}0.93\ $&  &  & $-0.31\ $ & $-0.13\ $&
$[ub]s-[sb]u\ $\\
$\Xi_b^\prime $ \eqref{B5n}\,
& $5.94\ $ & $5.99\ $ & $0.33\ $ & & $-0.12\ $ & $\phantom{-}0.93\ $ & $\phantom{-}0.045\ $ & $\phantom{-}0.099\ $& $\{us\}b\ $\\
$\Omega_b $ \eqref{B5k}\,
& $6.05\ $ & $6.12\ $ & $0.37\ $ & & $-0.12\ $ & $\phantom{-}0.91\ $ & $\phantom{-}0.049\ $ & $\phantom{-}0.12\ $& $\{ss\}b\ $\\
$\Xi_{cc}$ \eqref{B5o}\,
& $3.62\ $ & $3.72\ $ & $0.90\ $ & & $-0.32\ $ & $\phantom{-}0.26$ & $\phantom{-}0.097\ $ & $\phantom{-}0.057\ $& $[uc]c\ $\\
$\Xi_{cb}$ \eqref{B5q}\,
& $6.94^\ast\ $ & $7.10\ $ & $0.12\ $ & $-0.73\ $ &  &  & $\phantom{-}0.62\ $ & $-0.27\ $&
$[cb]u\ $\\  
$\Xi_{cb}^\prime$ \eqref{B5r}\,
& $6.97^\ast\ $ & $7.03\ $ & $0.90\ $ & & $\phantom{-}0.13\ $ & $\phantom{-}0.24\ $ & $-0.33\ $ & $\phantom{-}0.012\ $& $[uc]b+[ub]c\ $  \\  
$\Xi_{bb} $ \eqref{B5p}\,
& $\rule{-1.4ex}{0ex}10.14^\ast\ $ & $\rule{-1.4ex}{0ex}10.37\ $ & $0.87\ $ & & $-0.33\ $ & $\phantom{-}0.35\ $ & $\phantom{-}0.043\ $ & $\phantom{-}0.057\ $ & $[ub]b\ $\\ 
$\Omega_{cc} $ \eqref{B5s}\,
& $3.74^\ast\ $ & $3.90\ $ & $0.88\ $ & & $-0.33\ $ & $\phantom{-}0.30\ $ & $\phantom{-}0.15\ $ & $\phantom{-}0.074\ $& $[sc]c\ $\\
$\Omega_{cb} $ \eqref{B5u}\,
& $7.00^\ast\ $ & $7.22\ $ & $0.068\ $ & $-0.79\ $ & & & $\phantom{-}0.49\ $ & $-0.37\ $& $[cb]s\ $\\
$\Omega_{cb}^\prime $ \eqref{B5v}\,
& $7.03^\ast\ $ & $7.15\ $ & $0.85\ $ & & $-0.32\ $ & $-0.099\ $ & $\phantom{-}0.18\ $ & $\phantom{-}0.35\ $ & $[sc]b+[sb]c\ $\\
$\Omega_{bb} $ \eqref{B5t}\,
& $\rule{-1.4ex}{0ex}10.27^\ast\ $ & $\rule{-1.4ex}{0ex}10.48\ $ & $0.85\ $ & & $-0.36\ $ & $\phantom{-}0.37\ $ & $\phantom{-}0.065\ $ & $\phantom{-}0.085\ $ & $[sb]b\ $ \\
$\Omega_{ccb} $ \eqref{B5w}\,
& $8.01^\ast\ $ & $8.19\ $ & $0.39\ $ & & $-0.18\ $ & $\phantom{-}0.90\ $ & $\phantom{-}0.070\ $ & $\phantom{-}0.071\ $ & $\{cc\}b\ $\\
$\Omega_{cbb} $ \eqref{B5x}\,
& $\rule{-1.4ex}{0ex}11.20^\ast\ $ & $\rule{-1.4ex}{0ex}11.37\ $ & $0.81\ $ & & $-0.38\ $ & $\phantom{-}0.39\ $ & $\phantom{-}0.16\ $ & $\phantom{-}0.14\ $ & $[cb]b\ $\\\hline
\end{tabular*}
\end{center}
\end{table*}

Fig.\,\ref{figFaddeev} shows that the Faddeev kernels involve diquark breakup and reformation via exchange of a dressed-quark.  In order to present the most transparent analysis possible, we follow Refs.\,\cite{Roberts:2011cf, Chen:2012qr, Lu:2017cln} and introduce a simplification, \emph{viz}.\ in the Faddeev equation for a baryon of type $B$, the quark exchanged between the diquarks is represented as
\begin{equation}
S_f(k) \to \frac{g_B^2}{M_f}\,,
\label{staticexchange}
\end{equation}
where $f=l,s,c,b$ is the quark's flavour and $g_B$ is a coupling constant.  This is a variant of the  ``static approximation,'' which itself was introduced in Ref.\,\cite{Buck:1992wz}.  It has a marked impact on the Faddeev amplitudes, forcing them to be momentum-independent, just like the diquark Bethe-Salpeter amplitudes, but calculations reveal that it has little impact on the computed masses \cite{Xu:2015kta}.  We treat $g_{2\equiv 1/2^+}$, $g_{4\equiv 3/2^+}$ as parameters, choosing them below so as to obtain a desired mass for the lightest state in each $J^P$ sector.

\begin{figure*}[t]
\includegraphics[clip, width=0.99\textwidth]{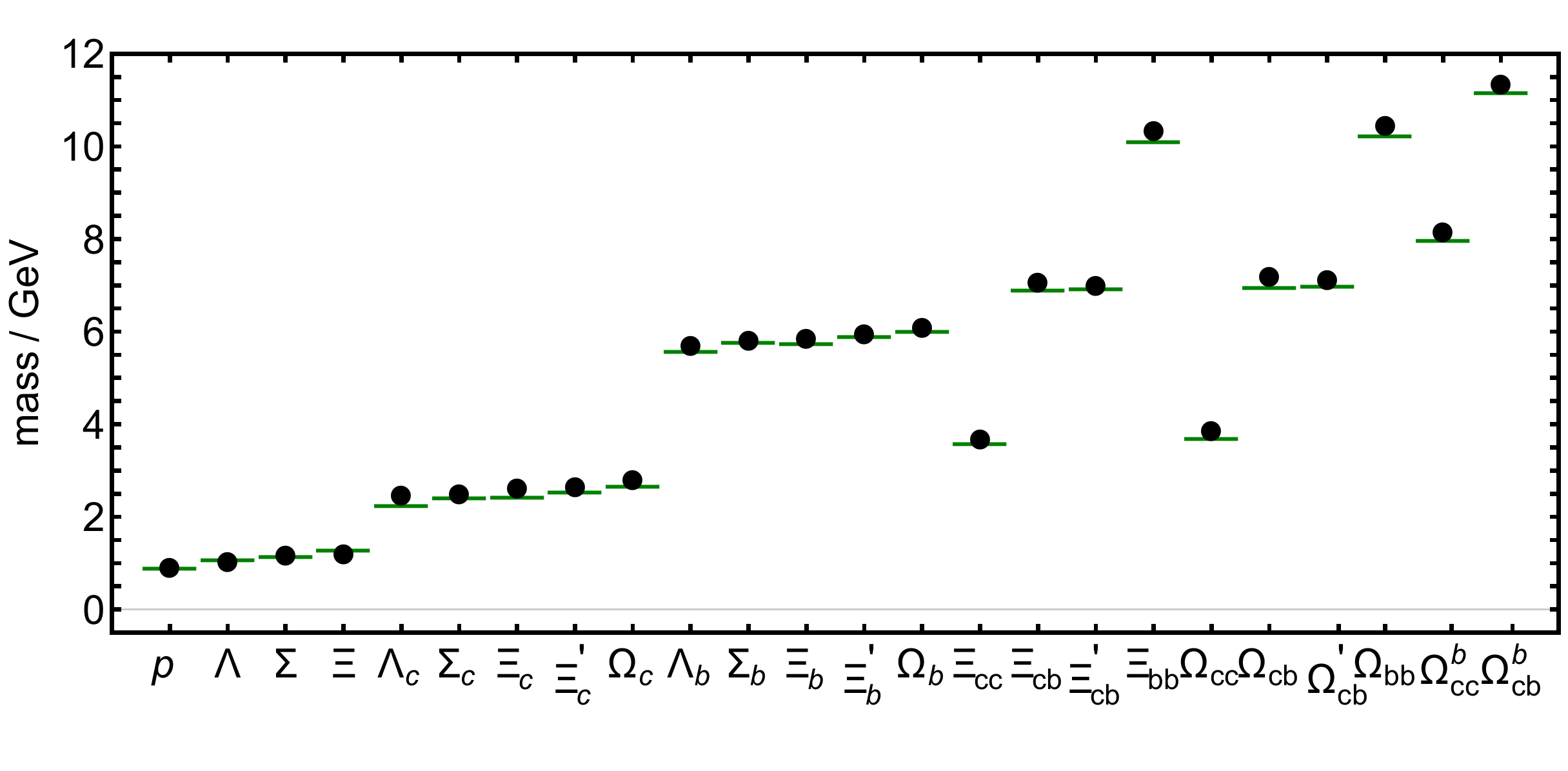}
\caption{\label{figResultsHalf}
Comparison between computed masses of ground-state flavour-$SU(5)$ $J^P=1/2^+$ baryons and either experiment  (first 15) \cite{Tanabashi:2018oca} or lQCD (last 9) \cite{Brown:2014ena, Mathur:2018epb}: our results -- (black) circles; and reference values  -- (green) bars.
}
\end{figure*}

The general forms of the matrices ${\cal S}^{\Psi}(\ell;P)$, ${\cal A}^{\Psi}_\nu(\ell;P)$ and ${\cal D}^{\Psi^{\wedge}}_{\nu\rho}(\ell;P)$ in Eqs.\,\eqref{calS}, \eqref{calA}, \eqref{calD}, respectively, which describe the momentum-space correlations between the quarks and diquarks in the baryons considered herein, are described in Refs.\,\cite{Oettel:1998bk, Cloet:2007pi}.  However, they simplify dramatically when Eq.\,\eqref{staticexchange} is used:
\begin{subequations}
\label{calSAcontact}
\begin{align}
 {\cal S}^{\Psi}(P) &= s^{\Psi}(P)\,\mathbf{I}_{\rm D} \,,\\
 {\cal A}^{\Psi}_\mu(P) &= a_1^\Psi(P)\,i\gamma_5\gamma_\mu + a_2^{\Psi}(P) \gamma_5 \hat P_\mu\,, \label{calAcontact}\\
{\cal D}^{\Psi^{\wedge}}_{\nu\rho}(\ell;P) & = a^{\Psi^\wedge}(P)\, \delta_{\nu\rho}\,.
\end{align}
\end{subequations}

\subsection{Faddeev Equations}
Specific examples of the Faddeev equations obtained from Fig.\,\ref{figFaddeev} for the various light-quark systems, along with their derivations, can be found in Ref.\,\cite{Chen:2012qr}.  Therefore, herein we only include equations for $\Xi_c^+$, $\Xi_c^{\prime +}$, $\Xi_c^{\ast +}$ because these three are sufficient to reconstruct all remaining Faddeev equations that are required to complete the spectrum calculations: one need only adjust the quark flavour labels.  The equations are given in Appendix~\ref{AppFEexplicit}.
It will be noted therefrom that we use the regularisation scheme explained in Sec.\,\ref{SecIS}, choosing $\Lambda_{\rm uv}$ to be the scale associated with the lightest diquark in the system, which is always the smallest value and hence the dominant regularising influence.

\subsection{Computed Masses and Amplitudes}
Eq.\,\eqref{staticexchange} introduces two parameters: $g_2$ in the $J^P=1/2^+$ sector and $g_4$ in the $J^P=3/2^+$ sector.  In this analysis, we fix these quantities by requiring that the relevant Faddeev equations return the empirical masses of the nucleon and $\Delta$-baryon:
\begin{equation}
\label{valueg2g4}
g_2 = 1.42\,,\quad g_4 = 1.96\,.
\end{equation}
It should be recalled that in choosing the couplings this way, various effects of resonant (meson cloud) contributions to hadron static properties are implicitly included \cite{Eichmann:2008ae}, and we capitalise on this feature herein.  However, some features are also omitted, \emph{e.g}.\ baryons computed using the kernel in Fig.\,\ref{figFaddeev} do not have widths, which are an essential physical consequence of meson-baryon final-state interactions (MB\,FSIs).  The operating conjecture for RL truncation is that the impact of MB\,FSIs on a baryon's Breit-Wigner mass is captured by the choice of interaction scale, even though a width is not generated.  This should be reasonable for states whose width is a small fraction of their mass.  In practice, the conjecture appears to be correct, at least for the ground-state $J^P=1/2^+$, $3/2^+$ systems \cite{Qin:2019hgk}.

\subsubsection{$J^P=1/2^+$}
\label{subsubHalf}
With every element now specified, it is straightforward to solve the algebraic Faddeev equations and obtain the masses and amplitudes for all ground-state flavour-$SU(5)$ $J^P=1/2^+, 3/2^+$ baryons.  Our results for the $1/2^+$ systems are listed in Table~\ref{ResultsHalf}.  We use a compact notation, based on the following observations.
For any given $J^P=1/2^+$ baryon, the spin-flavour structure is described by a three-entry column-vector.
The top row (${\rm r}_1$) reflects the strength of a scalar diquark in the baryon's Faddeev amplitude, the bottom row indicates that of an axial-vector diquark strength (${\rm r}_3$), and the middle entry (${\rm r}_2$) is either scalar or axial-vector, depending on the baryon.  Table~\ref{ResultsHalf} therefore includes a superscript to mark the row number in the appropriate line of Eqs.\,\eqref{uColumn}.
%

Our computed $J^P=1/2^+$ masses are compared with empirical/lQCD values in Fig.\,\ref{figResultsHalf}. The mean-absolute-relative-difference is $3.1(2.2)$\%.
This compares well with the fully-covariant three-body calculation described in Ref.\,\cite{Qin:2019hgk}, for which the analogous difference is $5.2(2.8)$\%.  Of course, that study is more sophisticated: it did not use a quark-diquark approximation; instead, the Faddeev equations were solved in a fully-consistent RL truncation.  Moreover, the baryon spectra in Ref.\,\cite{Qin:2019hgk} are \emph{ab initio} predictions, whereas we used a parameter ($g_2$ in Eqs.\,\eqref{staticexchange}, \eqref{valueg2g4}) to readjust the scale of the $J^P=1/2^+$ spectrum so that the proton mass matches experiment.  Notwithstanding these things, the agreement with Ref.\,\cite{Qin:2019hgk} indicates both that we have implemented a phenomenologically efficacious formulation of the contact interaction and that equal spacing rules must provide a good first approximation to our contact-interaction spectrum of $J^P=1/2^+$ baryons.

Table~\ref{ResultsHalf} also lists the contact-interaction quark-diquark Faddeev amplitudes for each baryon, which express structural characteristics of the associated bound-state.  The calculated results owe their values to both the fully-dynamical character of the diquark correlations, predicted by realistic analyses of the quark-quark scattering problem, and the nature of the Faddeev kernel, which ensures, through continual rearrangement, that every valence-quark participates actively in all diquark correlations to the fullest extent allowed by kinematics and symmetries.  Hence, the structure should be qualitatively independent of the quark-quark interaction's pointwise behaviour.  This expectation is supported by studies of ground-state $u,d,s$ octet and decuplet systems, in which the contact-interaction amplitudes have proven to be a reliable harbinger of the results obtained using more sophisticated kernels in Fig.\,\ref{figFaddeev} \cite{Chen:2019fzn}.

\begin{table}[t]
\caption{\label{ResultsThreeHalf}
Computed mass and Faddeev amplitude for each ground-state flavour-$SU(5)$ $J^P=3/2^+$ baryon: the last column highlights the baryon's dominant spin-flavour correlation.  Empirical mass values are taken from Ref.\,\cite{Tanabashi:2018oca}; where they are absent, lQCD results are listed \cite{Brown:2014ena, Mathur:2018epb} and indicated by ``$\ast$''.  The parenthesised label beside a baryon's name indicates the equation which specifies the associated Faddeev amplitude's spin-flavour vector.
(Mass in GeV.)
}
\begin{center}
\begin{tabular*}
{\hsize}
{
l@{\extracolsep{0ptplus1fil}}
|l@{\extracolsep{0ptplus1fil}}
l@{\extracolsep{0ptplus1fil}}
|c@{\extracolsep{0ptplus1fil}}
l@{\extracolsep{0ptplus1fil}}
l@{\extracolsep{0ptplus1fil}}
|c@{\extracolsep{0ptplus1fil}}}\hline
Baryon & $M^{e/l}\ $ & $M^{\rm CI}$ & $\ fff\ $ & \multicolumn{2}{c|}{other$\ $} & dom.\,corr.$\ $ \\\hline
$\Delta\ $ \eqref{B7a}\,
& $\ 1.23\ $ & $1.23\ $ & $1\ $ & & & $\{uu\}u\ $ \\
$\Sigma^\ast\ $ \eqref{B7b}\,
& $\ 1.38\ $ & $1.37\ $ &  & $ 0.60\ $ & $ 0.80\ $ & $\{us\}u\ $ \\
$\Xi^\ast\ $ \eqref{B7c}\,
& $\ 1.53\ $ & $1.49\ $ &  & $ 0.85\ $ & $ 0.52\ $ & $\{us\}s\ $ \\
$\Omega\ $ \eqref{B7d}\,
& $\ 1.67\ $ & $1.62\ $ & $ 1\ $ &  &  & $\{ss\}s\ $ \\
$\Sigma_c^\ast\ $ \eqref{B7e}\,
& $\ 2.52\ $ & $2.56\ $ &  & $ 0.64\ $ & $ 0.77\ $ & $\{uc\}u\ $ \\
$\Xi_c^\ast\ $ \eqref{B7g}\,
& $\ 2.66\ $ & $2.70\ $ &  & $ 0.63\ $ & $ 0.77\ $ & $\{uc\}s+\{sc\}u\ $ \\
$\Omega_c^\ast\ $ \eqref{B7f}\,
& $\ 2.77\ $ & $2.85\ $ &  & $ 0.62\ $ & $ 0.79\ $ & $\{sc\}s\ $ \\
$\Sigma_b^\ast\ $ \eqref{B7h}\,
& $\ 5.83\ $ & $5.84\ $ &  & $ 0.71\ $ & $ 0.71\ $ & $\{uu\}b \approx \{ub\}u\ $ \\
$\Xi_b^\ast\ $ \eqref{B7j}\,
& $\ 5.95\ $ & $5.98\ $ &  & $ 0.67\ $ & $ 0.74\ $ & $\{ub\}s+\{sb\}u\ $ \\
$\Omega_b^\ast\ $ \eqref{B7i}\,
& $\ 6.09^\ast\ $ & $6.11\ $ &  & $ 0.63\ $ & $ 0.78\ $ & $\{sb\}s\ $ \\
%
$\Xi_{cc}^\ast\ $ \eqref{B7k}\,
& $\ 3.69^\ast\ $ & $3.75\ $ &  & $ 0.98\ $ & $ 0.20\ $ & $\{uc\}c\ $ \\
$\Xi_{cb}^\ast\ $ \eqref{B7m}\,
& $\ 6.99^\ast\ $ & $7.07\ $ &  & $ 0.21\ $ & $ 0.98\ $ & $\{uc\}b+\{ub\}c\ $ \\
$\Xi_{bb}^\ast\ $ \eqref{B7l}\,
& $\ \rule{-1.4ex}{0ex}10.18^\ast\ $ & $\rule{-1.4ex}{0ex}10.34\ $ &  & $ 1.00\ $ & $ 0.078\ $ & $\{ub\}b\ $ \\
$\Omega_{cc}^\ast\ $ \eqref{B7n}\,
& $\ 3.82^\ast\ $ & $3.94\ $ &  & $ 0.96\ $ & $ 0.28\ $ & $\{sc\}c\ $ \\
$\Omega_{cb}^\ast\ $ \eqref{B7p}\,
& $\ 7.06^\ast\ $ & $7.21\ $ &  & $ 0.31\ $ & $ 0.95\ $ & $\{sc\}b+\{sb\}c\ $ \\
$\Omega_{bb}^\ast\ $ \eqref{B7o}\,
& $\ \rule{-1.4ex}{0ex}10.27^\ast\ $ & $\rule{-1.4ex}{0ex}10.47\ $ &  & $ 0.99\ $ & $ 0.12\ $ & $\{sb\}b\ $ \\ %
$\Omega_{ccc}^\ast\ $ \eqref{B7q}\,
& $\ 4.80^\ast\ $ & $5.00\ $ & $1\ $ &  &  & $\{cc\}c\ $ \\
$\Omega_{ccb}^\ast\ $ \eqref{B7r}\,
& $\ 8.04^\ast\ $ & $8.19\ $ &  & $0.65\ $ & $0.76\ $ & $\{cb\}c\ $ \\
$\Omega_{cbb}^\ast\ $ \eqref{B7s}\,
& $\ \rule{-1.4ex}{0ex}11.23^\ast\ $ & $\rule{-1.4ex}{0ex}11.38\ $ &  & $0.96\ $ & $0.28\ $ & $\{cb\}b\ $ \\
$\Omega_{bbb}^\ast\ $ \eqref{B7t}\,
& $\ \rule{-1.4ex}{0ex}14.37^\ast\ $ & $\rule{-1.4ex}{0ex}14.57\ $ & $1\ $ &  &  & $\{bb\}b\ $ \\\hline
\end{tabular*}
\end{center}
\end{table}

It is reasonable, therefore, to emphasise the following structural features of the ground-state flavour-$SU(5)$ $J^P=1/2^+$ baryons drawn from Table~\ref{ResultsHalf}.
\begin{description}
\item[$\mathpzc a$] The lightest allowed diquark correlation typically defines the most important component of a baryon's Faddeev amplitude, \emph{e.g}.\ the scalar $[uc]$ diquark dominates in the $\Xi_{cc}$ and the $[ub]$ is dominant in the $\Xi_{bb}$.  This remains true even if an axial-vector diquark is the lightest channel available, \emph{e.g}.\ $\{uu\}$ is dominant in the $\Sigma_c$.
\item[$\mathpzc b$] Dominance of the lightest diquark correlation can be overcome in flavour channels for which the spin-flavour structure of the bound-state and the quark-exchange character of the kernel in Fig.\,\ref{figFaddeev} lead dynamically to a preference for mixed-flavour correlations, \emph{e.g}.\ since the $[us]c$ combination in the $\Xi_c$ cannot reproduce itself, its strength may be fed into the $[uc]s-[sc]u$ correlation.  (See Appendix~\ref{FEexplicitXic}.)  The eventual outcome depends on the mass-scales of the kernel participants and how they affect rearrangment processes in the Faddeev kernel, \emph{e.g}.\ compare $\Xi_b$ with $\Xi_{cb}$.
\end{description}
%
%
%

These findings add to the argument against treatments of the three-body problem which assume that baryons can be described as effectively two-body in nature, \emph{e.g}.\ as being built from a constituent-quark and static/frozen constituent-diquark.  We verify that diquark correlations in QCD are essentially dynamical, and their breakup and reformation play a crucial role in defining baryon structure.  This is true for light-quark baryons, for which lQCD confirms that the spectrum possesses a richness that cannot be explained by a two-body model \cite{Edwards:2011jj}.  Furthermore, the consequences extend to baryons involving one or more heavy quarks, challenging both (\emph{i}) the treatment of singly-heavy baryons ($qq^\prime Q$, $q,q^\prime \in \{u,d,s\}$, $Q\in \{c,b\}$) as two-body light-diquark+heavy-quark ($qq^\prime+Q$) bound-states (see, e.g.\ Ref.\,\cite{Ebert:2011kk, Ferretti:2019zyh}) and (\emph{ii}) analyses of doubly-heavy baryons ($qQQ^\prime$) which assume such systems can be considered as two-body light-quark+heavy-diquark bound-states, $q + QQ^\prime$ (\emph{e.g}.\ Refs.\,\cite{Ebert:2002ig, Zhang:2008rt}).  These observations also have implications for few-body studies of the tetra- and penta-quark problems.

\subsubsection{$J^P=3/2^+$}
Our results for ground-state flavour-$SU(5)$ $J^P=3/2^+$ baryons are listed in Table~\ref{ResultsThreeHalf}.  In this case, the allowed spin-flavour combinations are simple.  Hence, we use only three columns to represent the Faddeev amplitude: rows with only one amplitude-entry describe baryons constituted from three valence-quarks of identical flavour, in which case the value is always unity; and rows with two such entries possess at least one unmatched valence-quark amongst the three.

\begin{figure*}[t]
\includegraphics[clip, width=0.99\textwidth]{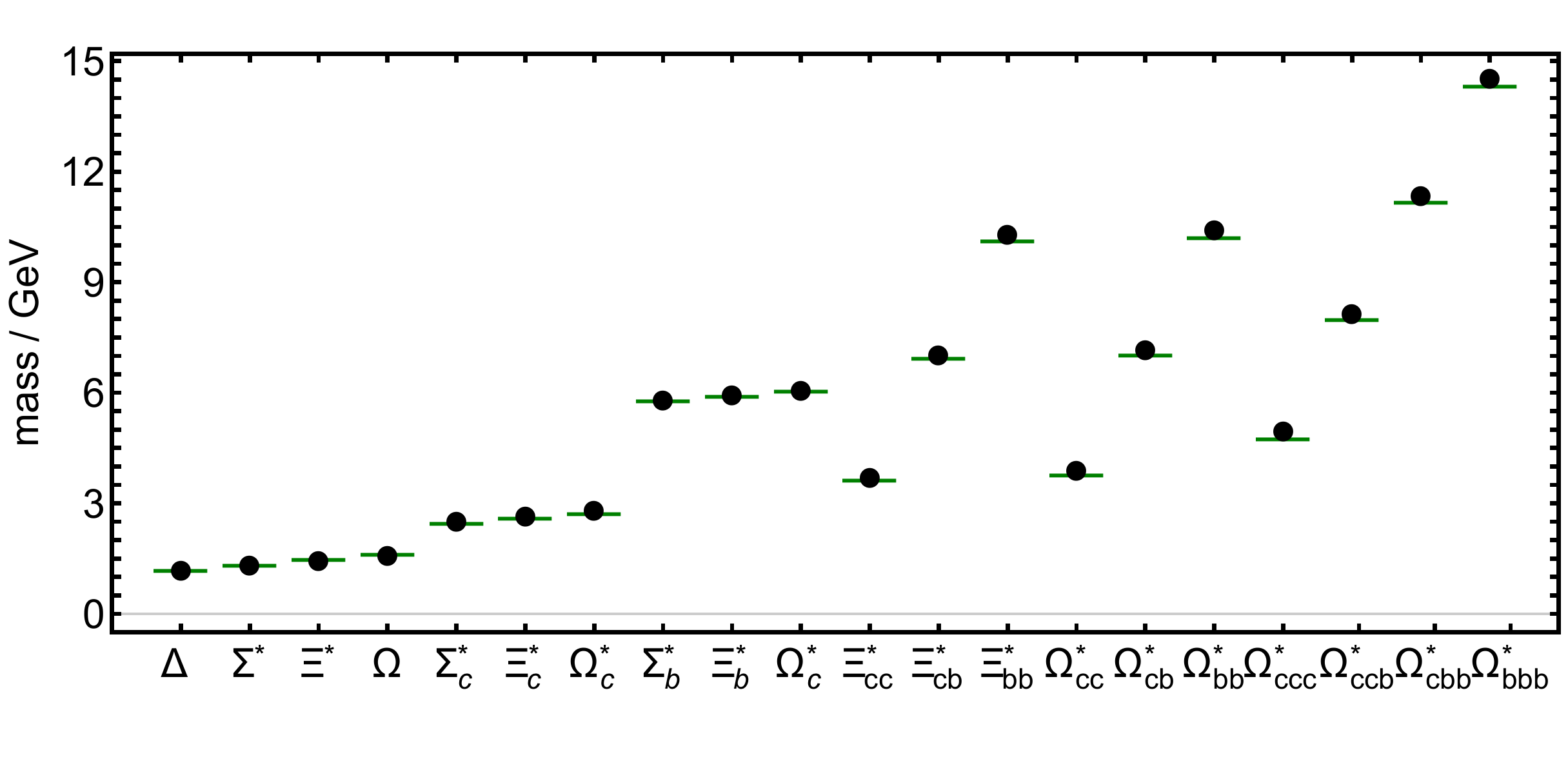}
\caption{\label{figResultsThreeHalf}
Comparison between computed masses of ground-state flavour-$SU(5)$ $J^P=3/2^+$ baryons and either experiment  (first 9) \cite{Tanabashi:2018oca} or lQCD (last 11) \cite{Brown:2014ena, Mathur:2018epb}: our results -- (black) circles; and reference values  -- (green) bars.
}
\end{figure*}

The computed $J^P=3/2^+$ masses are compared with empirical/lQCD values in Fig.\,\ref{figResultsThreeHalf}: the mean-absolute-relative-difference is $1.8(1.0)$\%.
Once again, this compares well with the fully-covariant three-body calculation described in Ref.\,\cite{Qin:2019hgk}, for which the analogous difference is $2.6(1.6)$\%.  Of course, here we used a parameter ($g_4$ in Eqs.\,\eqref{staticexchange}, \eqref{valueg2g4}) to readjust the scale of the $J^P=3/2^+$ spectrum so that the $\Delta$-baryons's mass matches experiment.  Nevertheless, as before, the agreement with Ref.\,\cite{Qin:2019hgk} highlights both the utility of our formulation of the contact interaction and the validity of equal spacing rules as a first approximation to the spectrum of $J^P=3/2^+$ baryons.

Table~\ref{ResultsThreeHalf} also lists the contact-interaction quark-diquark Faddeev amplitudes for each baryon, which express structural characteristics of the associated bound-state.  As in Sec.\,\ref{subsubHalf}, the calculated results owe their values to both the fully-dynamical character of the diquark correlations and the nature of the Faddeev kernel, which together ensure a continual shuffling of each dressed-quark into and out of diquark correlations.  The structure of this kernel for $J^P=3/2^+$ baryons is exemplified by that sketched for the $\Xi_c^{\ast}$ in Appendix~\ref{AppXicstar}.  Its form makes clear that in all cases involving more than one quark flavour, the diquark combination with maximal flavour shuffling is favoured because it is fed by twice as many exchange processes as the less-mixed correlation.
The $\Sigma_b^\ast$ is the only case in which this outcome is avoided as a consequence of the enormous mass-splitting between the $u$- and $b$-quarks, which dramatically suppresses the kernel contribution
\begin{equation}
u\{bu\} \stackrel{\mbox{$b$-exchange}}{\to} \{ub\} u\,,
\end{equation}
thereby producing a roughly-symmetric, anti-diagonal kernel and hence a solution with near equality between the two possible diquark combinations.

\section{Summary and Outlook}
\label{epilogue}
We employed a confining, symmetry-preserving treatment of a vector$\,\times\,$vector contact-interaction to calculate spectra of ground-state pseudoscalar and vector $(f\bar g)$ mesons, scalar and axial-vector $(fg)$ diquarks, and $J^P=1/2^+, 3/2^+$ $(fgh)$ baryons, where $f,g,h \in \{u,d,s,c,b\}$.  A physically-motivated refinement of earlier formulations, based on feedback between pseudoscalar-meson masses, $m_{0^-}$, and leptonic decay constants, leading to a logarithmic evolution of the interaction strength with $m_{0^-}$, was necessary to extend applicability of the contact-interaction to systems involving heavy quarks (Sec.\,\ref{SecIS}).  The calculated meson spectrum agrees well with experiment (Sec.\,\ref{SecMeson}): the mean-relative-difference for 15 states is 5(5)\%; and for these same states, the leptonic decay constants are reproduced with an accuracy of 18(9)\%.

Diquark masses and correlation strengths are required as input to our baryon Faddeev equations; and a straightforward relationship between quark-quark and quark-antiquark scattering means that these quantities are readily obtained via a single, simple modification of the meson Bethe-Salpeter equations.  It follows that for every $J^{-P}$ multiplet of degenerate mesons there is an associated $J^P$ diquark multiplet (with, perhaps, just one member).  The level ordering of the $J^{P}$ diquark correlations matches that of the $J^{-P}$ meson partners, with the meson masses bounding the partner diquark masses from below (Sec.\,\ref{SecDiquarks}): ignoring diquarks partnered with Nambu-Goldstone modes, the splitting is $0.13(6)\,$GeV.

In proceeding to solve the baryon Faddeev equations, we used a static approximation to the quark exchange kernel.  This produces Faddeev amplitudes that are momentum-independent and thereby ensures a level of consistency with the two-body bound-state amplitudes generated by the contact-interaction.  Our implementation of the static approximation introduced two parameters.  Choosing them to ensure that the masses of the lightest $J^P=1/2^+$, $3/2^+$ baryons agreed with experiment, we arrived at predictions for the masses of 42 ground-state flavour-$SU(5)$ $J^P=1/2^+, 3/2^+$ baryons which compare well with measurements (22 states known) or lQCD results (20 states): the mean-absolute-relative difference is $2.5(1.1)$\% (Figs.\,\ref{figResultsHalf}, \ref{figResultsThreeHalf}).

The framework employed herein is distinguished by its simplicity, with analyses and calculations being largely algebraic.  It is also marked by its effectiveness.  There are six parameters: four used to define the interaction and its scale dependence through analysis of $\pi$, $K$, $\eta_c$, $\eta_b$ properties; and two introduced to boost attraction in the baryon Faddeev equations.  Thus defined, the contact-interaction delivers predictions for 72 distinct quantities; and when one compares with independent determinations, the mean-absolute-relative difference is $6(8)$\%.

Such quantitative success suggests that serious consideration should be given to the qualitative conclusions supported by our analysis.  Of primary significance is the demonstration that diquark correlations play an important role in all baryons studied.  Crucially, these diquark correlations are not frozen degrees-of-freedom within a baryon.  Instead, they are \emph{dynamically composite}, an aspect which entails that it is the lightest allowed diquark correlation which typically defines the most important component of a baryon's Faddeev amplitude.  This remains true even if the (\emph{bad}) axial-vector diquark is the lightest channel available.  The dominance of the lightest correlation can only be overwhelmed if the baryon's spin-flavour structure is such that the dynamical diquark breakup and reformation processes driven by the Faddeev kernel lead to a preference for mixed-flavour correlations.

It is straightforward to generalise our analysis to negative-parity baryon ground-states and we anticipate that the picture drawn therefrom would be qualitatively sound \cite{Lu:2017cln, Chen:2017pse}.  An extension to the first positive-parity excitations of the baryons considered herein is also possible, but one is forced into contrivance when trying to describe baryons with zeros in their Faddeev amplitudes using an interaction that favours momentum-independent solutions \cite{Lu:2017cln, Chen:2017pse, Segovia:2015hra, Burkert:2017djo}.  Additionally, given that novel gluon-quark/gluon-antiquark correlations might play a part in explaining hybrid mesons \cite{Xu:2018cor}, a new and worthwhile direction would be to adapt the framework described herein to the challenge of understanding tetra- and penta-quark states, in which dynamical diquark-correlations may also be expected to play a material role.

In all cases, too, it is sensible to check conclusions reached using the contact-interaction against those obtained with realistic momentum-dependent kernels for the three-valence-body bound-state problem.  As already noted, this task has been undertaken for many flavour-$SU(3)$ baryons using the quark-diquark approximation \cite{Chen:2017pse, Chen:2019fzn}, although the diquark-content predictions have not been fully tested.  Such analyses should be completed and also extended to flavour-$SU(5)$.

Direct solutions of the three-body problem, eschewing the quark-diquark approximation and using a well-constrained rainbow-ladder kernel, have been employed to produce the spectrum of ground-state flavour-$SU(5)$ $(fgh)$ baryons, their first positive-parity excitations and parity partners \cite{Qin:2019hgk}.  The solutions could profitably be used to explore for signals indicating the appearance of diquark correlations.  It is also reasonable to pursue extensions of these direct studies with kernels improved by incorporating nonperturbative effects of dynamical chiral symmetry breaking, which are potentially important to the emergence of strong diquark correlations.  Such efforts are likely to benefit from the use of high-performance computing.

\begin{acknowledgments}


We are grateful for constructive comments and encouragement from \mbox{Z.-F.~Cui}, R.~Gothe, V.~Mokeev, S.\,M.~Schmidt, F.~Wang and \mbox{H.-S.~Zong};
and for the hospitality and support of the University of Huelva, Huelva - Spain, and the University of Pablo de Olavide, Seville - Spain, during the ``4th Workshop on Nonperturbative QCD'' at the University of Pablo de Olavide, 6-9 November 2018.
Work supported by:
National Natural Science Foundation of China, under grant no.\ 11805097;
Jiangsu Province Natural Science Foundation, under grant no.\ BK20180323;
Jiangsu Province \emph{Hundred Talents Plan for Professionals};
Conselho Nacional de Desenvolvimento Cient{\'{\i}}fico e Tecnol{\'o}gico - CNPq, Grant Nos.\,305894/2009-9, 464898/2014-5 (INCT F{\'{\i}}sica Nuclear e Aplica{\c{c}}{\~o}es);
Funda\c{c}\~ao de Amparo \`a Pesquisa do Estado de S\~ao Paulo - FAPESP Grant Nos.\,2013/01907-0,
2015/21550-4;
U.S.\ Department of Energy, Office of Science, Office of Nuclear Physics, under contract no.\,DE-AC02-06CH11357;
and Ministerio de Economia Industria y Competitividad (MINECO), under grant no.\ FPA2017-86380-P.
\end{acknowledgments}


\appendix
\setcounter{equation}{0}
\setcounter{figure}{0}
\setcounter{table}{0}
\renewcommand{\theequation}{\Alph{section}.\arabic{equation}}
\renewcommand{\thetable}{\Alph{section}.\arabic{table}}
\renewcommand{\thefigure}{\Alph{section}.\arabic{figure}}

\section{Collected Formulae}
\label{AppFormulae}
Eq.\,\eqref{gapactual} is the first of a number of integrals appearing in our analysis whose regularised values are expressed in terms of incomplete gamma-functions.  We gather the expressions here $(n=0,1,2)$:
\begin{equation}
\label{Cndef}
{\cal C}_n^{\rm iu}(\sigma) = \sigma \overline{\cal C}_n^{\rm iu}(\sigma)\,,
\end{equation}
with
\begin{subequations}
\label{C012def}
\begin{align}
\overline{\cal C}_0^{\rm iu}(\sigma) &
= \Gamma(-1,\sigma \tau_{\rm ir}^2) - \Gamma(-1,\sigma \tau_{\rm uv}^2)\,,\\
\overline{\cal C}_1^{\rm iu}(\sigma) & = - \frac{d}{d\sigma} {\cal C}_0^{\rm iu}(\sigma)
= \Gamma(0,\sigma \tau_{\rm ir}^2) - \Gamma(0,\sigma \tau_{\rm uv}^2)\,,\\
\overline{\cal C}_2^{\rm iu}(\sigma) & = \sigma \frac{d^2}{d\sigma^2} {\cal C}_0^{\rm iu}(\sigma)
= \Gamma(1,\sigma \tau_{\rm ir}^2) - \Gamma(1,\sigma \tau_{\rm uv}^2)\,,
\end{align}
\end{subequations}
or, more simply still,
\begin{equation}
\overline{\cal C}_n^{\rm iu}(\sigma) =
\Gamma(n-1,\sigma \tau_{\rm ir}^2) - \Gamma(n-1,\sigma \tau_{\rm uv}^2)\,.
\end{equation}

These expressions are used, \emph{e.g}.\ to express the Bethe-Salpeter kernel for pseudoscalar mesons in Eq.\,\eqref{bsefinalE}:
{\allowdisplaybreaks
\begin{subequations}
\label{fgKernel}
\begin{eqnarray}
\nonumber
{\cal K}_{EE}^{0^-} &=&
\int_0^1d\alpha \bigg\{
{\cal C}_0^{\rm iu}(\omega_{f g}( \alpha, Q^2))  \\
&&+ \bigg[ M_f M_{g}-\alpha \hat\alpha Q^2 - \omega_{f g}( \alpha, Q^2)\bigg]\nonumber \\
&&
\quad \times
\overline{\cal C}^{\rm iu}_1(\omega_{f g}(\alpha, Q^2))\bigg\},\\
\nonumber
{\cal K}_{EF}^{0^-} &=& \frac{Q^2}{2 M_{f g}} \int_0^1d\alpha\, \bigg[\hat \alpha M_f+\alpha M_{g}\bigg]\\
&& \quad \times \overline{\cal C}^{\rm iu}_1(\omega_{f g}(\alpha, Q^2)),\\
{\cal K}_{FE}^{0^-} &=& \frac{2 M_{f g}^2}{Q^2} {\cal K}_{EF}^{0^-} ,\\
\nonumber
{\cal K}_{FF}^{0^-} &=& - \frac{1}{2} \int_0^1d\alpha\, \bigg[ M_f M_{g}+\hat\alpha M_f^2+\alpha M_{g}^2\bigg]\\
&& \quad \times \overline{\cal C}^{\rm iu}_1(\omega_{f g}(\alpha, Q^2))\,.
\end{eqnarray}
\end{subequations}}

It is worth reiterating here that Eq.\,\eqref{bsefinalE} is an eigenvalue problem that has a solution for $Q^2=-m_{0^-}^2$, at which point the eigenvector is the meson's Bethe-Salpeter amplitude.  In the computation of observables one must employ the canonically normalised amplitude, \emph{viz}.\ the amplitude rescaled such that
\begin{equation}
\label{normcan}
1=\left. \frac{d}{d Q^2}\Pi_{0^-}(Z,Q)\right|_{Z=Q},
\end{equation}
where
\begin{eqnarray}
\nonumber \Pi_{0^-}(Z,Q) &=& 6 {\rm tr}_{\rm D} \!\! \int\! \frac{d^4t}{(2\pi)^4} \Gamma_{0^-}(-Z)\\
&& \quad \times
 S_f(t+Q) \, \Gamma_{0^-}(Z)\, S_g(t)\,. \label{normcan2}
\end{eqnarray}

\section{Elements in the Faddeev Equations}
\label{AppFE}
It is useful to define a collection of flavour matrices to assist with explicating the structure of the diquark pieces in Eq.\,(\ref{Psi}).  To do so, we realise the fundamental representation of flavour-$SU(5)$ as follows:
\begin{equation}
\label{flavourarrays}
\begin{array}{ccc}
u = \left[\begin{array}{ccccc}
                    1 \\
                    0\\
                    0\\
                    0 \\
                    0
                    \end{array}\right],
&
d = \left[\begin{array}{ccccc}
                    0 \\
                    1\\
                    0\\
                    0 \\
                    0
                    \end{array}\right],
&
s = \left[\begin{array}{ccccc}
                    0 \\
                    0\\
                    1\\
                    0 \\
                    0
                    \end{array}\right],\\
c = \left[\begin{array}{ccccc}
                    0 \\
                    0\\
                    0\\
                    1 \\
                    0
                    \end{array}\right],
&
b = \left[\begin{array}{ccccc}
                    0 \\
                    0\\
                    0\\
                    0 \\
                    1
                    \end{array}\right];
                    &
\end{array}
\end{equation}
These states can be combined into diquark correlations using the following $5\times 5$ matrices, where here we distinguish between $u$- and $d$-quarks:
{\allowdisplaybreaks
\begin{subequations}
\label{flavourarrays1}
\begin{align}
{\tt t}^{1=[ud]} = \left[\begin{array}{ccccc}
                    0 & 1 & 0&0&0 \\
                    -1 & 0 & 0&0&0 \\
                    0 & 0 & 0&0&0\\
                    0 & 0 & 0&0&0\\
                    0 & 0 & 0&0&0
                    \end{array}\right], \nonumber\\
{\tt t}^{2=[us]} = \left[\begin{array}{ccccc}
                    0 & 0 & 1&0&0 \\
                    0 & 0 & 0&0&0 \\
                    -1 & 0 & 0&0&0\\
                    0 & 0 & 0&0&0\\
                    0 & 0 & 0&0&0
                    \end{array}\right],\\
{\tt t}^{3=[uc]} = \left[\begin{array}{ccccc}
                    0 & 0 & 0&1&0 \\
                    0 & 0 & 0&0&0 \\
                    0 & 0 & 0&0&0\\
                    -1 & 0 & 0&0&0\\
                    0 & 0 & 0&0&0
                    \end{array}\right], \nonumber
\end{align}
\end{subequations}}
\hspace*{-0.7\parindent}with ${\tt t}^{4=[ub]}$, ${\tt t}^{5=[ds]}$, ${\tt t}^{6=[dc]}$, ${\tt t}^{7=[db]}$, ${\tt t}^{8=[sc]}$, ${\tt t}^{9=[sb]}$, ${\tt t}^{10=[cb]}$ obvious by analogy; and
{\allowdisplaybreaks
\begin{subequations}
\label{flavourarrays2}
\begin{align}
{\tt t}^{11=\{uu\}} & = \left[\begin{array}{ccccc}
                    \sqrt{2} &0 & 0&0&0 \\
                    0 & 0 & 0&0&0 \\
                    0 & 0 & 0&0&0\\
                    0 & 0 & 0&0&0\\
                    0 & 0 & 0&0&0
                    \end{array}\right], \nonumber \\
{\tt t}^{12=\{ud\}} & = \left[\begin{array}{ccccc}
                    0 & 1 & 0&0&0 \\
                    1 & 0 & 0&0&0 \\
                    0 & 0 & 0&0&0\\
                    0 & 0 & 0&0&0\\
                    0 & 0 & 0&0&0
                    \end{array}\right],\\
{\tt t}^{13=\{us\}} & = \left[\begin{array}{ccccc}
                    0 & 0 & 1&0&0 \\
                    0 & 0 & 0&0&0 \\
                    1 & 0 & 0&0&0\\
                    0 & 0 & 0&0&0\\
                    0 & 0 & 0&0&0
                    \end{array}\right], \nonumber
\end{align}
\end{subequations}
}
\hspace*{-0.5\parindent}with ${\tt t}^{14=\{uc\}}$, ${\tt t}^{15=\{ub\}}$, ${\tt t}^{16=\{dd\}}$, ${\tt t}^{17=\{ds\}}$, ${\tt t}^{18=\{dc\}}$, ${\tt t}^{19=\{db\}}$, ${\tt t}^{20=\{ss\}}$, ${\tt t}^{21=\{sc\}}$, ${\tt t}^{22=\{sb\}}$, ${\tt t}^{23=\{cc\}}$, ${\tt t}^{24=\{cb\}}$, ${\tt t}^{25=\{bb\}}$ also clear by analogy.

Now with the dressed-quark propagators collected into a diagonal flavour-matrix:
\begin{align}
{\mathsf S}(p)= \left[
\begin{array}{ccccc}
S_u(p) & 0 & 0 & 0 & 0 \\
0 & S_d(p)& 0 & 0 & 0  \\
0 & 0 & S_s(p) & 0 & 0 \\
0 & 0 & 0 & S_c(p) & 0 \\
0 & 0 & 0 & 0 & S_b(p)
\end{array}
\right],
\end{align}
then, \emph{e.g}.\ the $0^+$ $[ub]$ diquark is represented by the flavour-matrix ${\tt t}^{4=[ub]}$ and its contribution to a Bethe-Salpeter or Faddeev kernel is obtained through the matrix product ${\mathsf S}(p_1) t^4 {\mathsf S}^{\rm T}(p_2)$\,, where $(\cdot)^{\rm T}$ indicates matrix transpose.  It follows that the baryons considered herein have the spin-flavour structures listed below.\footnote{We capitalise on isospin symmetry, \emph{i.e}.\ all diquarks and baryons in an isospin multiplet are degenerate; and, hence, solve only for the baryon within a given isospin multiplet whose flavour structure is simplest.}
\subsection{$1/2^+$ Spin-Flavour Combinations}
In the isospin-symmetry limit, using the quark-diquark approximation, the Faddeev equation in Fig.\,\ref{figFaddeev} supports the following $24$ distinct flavour-$SU(5)$ combinations for $1/2^+$ baryons.
{\allowdisplaybreaks
\begin{subequations}
\label{uColumn}
\begin{align}
u_p & = \left[
\begin{array}{c}
[ud] u \\
\{uu\} d \\
\{ud\} u
\end{array} \right]
\leftrightarrow
\left[
\begin{array}{l}
{\mathpzc s}^1\\
{\mathpzc a}^{11}\\
{\mathpzc a}^{12}
\end{array}
\right],
\label{B5a}\\
u_\Lambda& = \frac{1}{\surd 2}
\left[
\begin{array}{c}
\surd 2 [ud] s \\
\, [us]d - [ds] u \\
\{us\} d - \{d s\} u
\end{array}
\right]
\leftrightarrow
\left[
\begin{array}{l}
{\mathpzc s}^1\\
{\mathpzc s}^{[2,5]}\\
{\mathpzc a}^{[13,17]}
\end{array}
\right], \label{B5d}\\
u_{\Sigma^+} & =
\left[ \begin{array}{c}
[us] u \\
\{uu\} s \\
\{us\} u
\end{array} \right]
\leftrightarrow
\left[
\begin{array}{l}
{\mathpzc s}^2\\
{\mathpzc a}^{11}\\
{\mathpzc a}^{13}
\end{array}
\right], \label{B5b}\\
u_{\Xi^0} & =
\left[
\begin{array}{c}
[us] s \\
\{us\} s \\
\{ss\} u
\end{array}\right]
\leftrightarrow
\left[
\begin{array}{l}
{\mathpzc s}^2\\
{\mathpzc a}^{13}\\
{\mathpzc a}^{20}
\end{array}
\right], \label{B5c}\\
%
u_{\Lambda_{c}^{+}}& =\frac{1}{\sqrt{2}}
\left[
\begin{array}{c}
\sqrt{2} [ud] c \\
\, [uc]d - [dc] u \\
\{uc\} d - \{d c\} u
\end{array}\right]
\leftrightarrow
\left[
\begin{array}{l}
{\mathpzc s}^1\\
{\mathpzc s}^{[3,6]}\\
{\mathpzc a}^{[14,18]}
\end{array}
\right], \label{B5g}\\
u_{\Sigma_{c}^{++}} & = \left[
\begin{array}{c}
[uc] u \\
\{uu\} c \\
\{uc\} u
\end{array} \right]
\leftrightarrow
\left[
\begin{array}{l}
{\mathpzc s}^3\\
{\mathpzc a}^{11}\\
{\mathpzc a}^{14}
\end{array}
\right], \label{B5e}\\
u_{\Xi_{c}^{+}}&= \frac{1}{\sqrt{2}}
\left[
\begin{array}{c}
\sqrt{2} [us] c \\
\, [uc]s - [sc] u \\
\{uc\} s - \{sc\} u
\end{array}
\right]
\leftrightarrow
\left[
\begin{array}{l}
{\mathpzc s}^2\\
{\mathpzc s}^{[3,8]}\\
{\mathpzc a}^{[14,21]}
\end{array}
\right],
\label{Xic} \\
u_{\Xi_{c}^{\prime+}}& = \frac{1}{\sqrt{2}}
\left[
\begin{array}{c}
\, [uc]s + [sc] u \\
\sqrt{2} \{us\} c \\
\{uc\} s + \{sc\}u
\end{array}
\right]
\leftrightarrow
\left[
\begin{array}{l}
{\mathpzc s}^{\{3,8\}}\\
{\mathpzc a}^{13}\\
{\mathpzc a}^{\{14,21\}}
\end{array}
\right],
\label{Xicprime}\\
u_{\Omega_{c}^{0}}& =
\left[ \begin{array}{c}
[sc] s \\
\{ss\} c \\
\{sc\} s
\end{array} \right],
\leftrightarrow
\left[
\begin{array}{l}
{\mathpzc s}^8\\
{\mathpzc a}^{20}\\
{\mathpzc a}^{21}
\end{array}
\right], \label{B5f}\\
%
u_{\Lambda_{b}^{0}} & =\frac{1}{\sqrt{2}}
\left[
\begin{array}{c}
\sqrt{2} [ud] b \\
\, [ub]d - [db] u \\
\{ub\} d - \{db\} u
\end{array}\right]
\leftrightarrow
\left[
\begin{array}{l}
{\mathpzc s}^1\\
{\mathpzc s}^{[4,7]}\\
{\mathpzc a}^{[15,19]}
\end{array}\right],
\label{B5l}\\
u_{\Sigma_{b}^{+}} & = \left[
\begin{array}{c}
[ub] u \\
\{uu\} b \\
\{ub\} u
\end{array} \right]
\leftrightarrow
\left[
\begin{array}{l}
{\mathpzc s}^4\\
{\mathpzc a}^{11}\\
{\mathpzc a}^{15}
\end{array}\right],
\label{B5j}\\
u_{\Xi_{b}^{0}} & = \frac{1}{\sqrt{2}}
\left[
\begin{array}{c}
\sqrt{2} [us] b \\
\, [ub]s - [sb] u \\
\{ub\} s - \{sb\} u
\end{array}
\right]
\leftrightarrow
\left[
\begin{array}{l}
{\mathpzc s}^2\\
{\mathpzc s}^{[4,9]}\\
{\mathpzc a}^{[15,22]}
\end{array}\right],
\label{B5m}\\
u_{\Xi_{b}^{\prime0}} & = \frac{1}{\sqrt{2}}
\left[
\begin{array}{c}
\, [ub]s + [sb] u \\
\sqrt{2} \{us\} b \\
\{ub\} s + \{sb\}u
\end{array}
\right]
\leftrightarrow
\left[
\begin{array}{l}
{\mathpzc s}^{\{4,9\}}\\
{\mathpzc a}^{13}\\
{\mathpzc a}^{\{15,22\}}
\end{array}
\right], \label{B5n}\\
u_{\Omega_{b}^{-}} & =
\left[ \begin{array}{c}
[sb] s \\
\{ss\} b \\
\{sb\} s
\end{array} \right]
\leftrightarrow
\left[
\begin{array}{l}
{\mathpzc s}^9\\
{\mathpzc a}^{20}\\
{\mathpzc a}^{22}
\end{array}\right],
\label{B5k}\\
u_{\Xi_{cc}^{++}} & = \left[
\begin{array}{c}
[uc] c \\
\{uc\} c \\
\{cc\} u
\end{array} \right]
\leftrightarrow
\left[
\begin{array}{l}
{\mathpzc s}^3\\
{\mathpzc a}^{14}\\
{\mathpzc a}^{23}
\end{array}\right],
\label{B5o}\\
u_{\Xi_{cb}^{+}} & = \frac{1}{\sqrt{2}}
\left[
\begin{array}{c}
\, [uc]b - [ub]c \\
\sqrt{2} [cb]u\\
\{uc\}b - \{ub\} c
\end{array}
\right]
\leftrightarrow
\left[
\begin{array}{l}
{\mathpzc s}^{[3,4]}\\
{\mathpzc s}^{10}\\
{\mathpzc a}^{[14,15]}
\end{array}\right],
\label{B5q}\\
u_{\Xi_{cb}^{\prime+}} & = \frac{1}{\sqrt{2}}
\left[
\begin{array}{c}
\, [uc]b + [ub]c \\
\{uc\}b + \{ub\} c \\
\sqrt{2} \{cb\}u\\
\end{array}
\right]
\leftrightarrow
\left[
\begin{array}{l}
{\mathpzc s}^{\{3,4\}}\\
{\mathpzc a}^{\{14,15\}}\\
{\mathpzc a}^{24}
\end{array}\right],
\label{B5r}\\
u_{\Xi_{bb}^{0}} & =
\left[ \begin{array}{c}
[ub]b \\
\{ub\}b \\
\{bb\}u
\end{array} \right]
\leftrightarrow
\left[
\begin{array}{l}
{\mathpzc s}^4\\
{\mathpzc a}^{15}\\
{\mathpzc a}^{25}
\end{array}\right],
\label{B5p}\\
u_{\Omega_{cc}^{+}} & = \left[
\begin{array}{c}
[sc] c \\
\{sc\} c \\
\{cc\} s
\end{array} \right]
\leftrightarrow
\left[
\begin{array}{l}
{\mathpzc s}^8\\
{\mathpzc a}^{21}\\
{\mathpzc a}^{23}
\end{array}\right],
\label{B5s}\\
u_{\Omega_{cb}^{0}} & = \frac{1}{\sqrt{2}}
\left[
\begin{array}{c}
\, [sc]b - [sb]c \\
\sqrt{2} [cb]s\\
\{sc\}b - \{sb\} c
\end{array}
\right]
\leftrightarrow
\left[
\begin{array}{l}
{\mathpzc s}^{[8,9]}\\
{\mathpzc s}^{10}\\
{\mathpzc a}^{[21,22]}
\end{array}\right],
\label{B5u}\\
u_{\Omega_{cb}^{\prime 0}} & = \frac{1}{\sqrt{2}}
\left[
\begin{array}{c}
\, [sc]b + [sb]c \\
\{sc\}b + \{sb\} c \\
\sqrt{2} \{cb\}s\\
\end{array}
\right]
\leftrightarrow
\left[
\begin{array}{l}
{\mathpzc s}^{\{8,9\}}\\
{\mathpzc a}^{\{21,22\}}\\
{\mathpzc a}^{24}
\end{array}\right],
\label{B5v}\\
u_{\Omega_{bb}^{-}} & =
\left[ \begin{array}{c}
[sb]b \\
\{sb\}b \\
\{bb\}s
\end{array} \right]
\leftrightarrow
\left[
\begin{array}{l}
{\mathpzc s}^9\\
{\mathpzc a}^{22}\\
{\mathpzc a}^{25}
\end{array}\right],
\label{B5t}\\
u_{\Omega_{ccb}^{+}} & = \left[
\begin{array}{c}
[cb] c \\
\{cc\}b \\
\{cb\} c
\end{array} \right]
\leftrightarrow
\left[
\begin{array}{l}
{\mathpzc s}^{10}\\
{\mathpzc a}^{23}\\
{\mathpzc a}^{24}
\end{array}\right],
\label{B5w}\\
u_{\Omega_{cbb}^{0}} & =
\left[
\begin{array}{c}
[cb]b \\
\{cb\}b \\
\{bb\}c
\end{array} \right]
\leftrightarrow
\left[
\begin{array}{l}
{\mathpzc s}^{10}\\
{\mathpzc a}^{24}\\
{\mathpzc a}^{25}
\end{array}\right].
\label{B5x}
\end{align}
\end{subequations}}

Sec.\,\ref{SecDiquarks} reveals that the antisymmetric combination of any two quark flavours is always lighter than the symmetric combination.  Consequently, in the absence of additional contributions to the Faddeev kernel, one should expect
{\allowdisplaybreaks
\begin{subequations}
\begin{align}
M_{\Lambda^0} & < M_{\Sigma^0} , \\
M_{\Xi_c} & < M_{\Xi_c^\prime} , \\
M_{\Xi_b} & < M_{\Xi_b^\prime} .
\end{align}
\end{subequations}}
\hspace*{-0.5ex}Moreover, the level ordering pattern should reverse for $(M_{\Xi_{cb}},M_{\Xi_{cb}^\prime})$ and $(M_{\Omega_{cb}},M_{\Omega_{cb}^\prime})$ because
the $0^+$ $[cb]$ diquark correlation is $\approx 1\,$GeV more massive than the $[l,b]$, $[s,b]$ diquarks, %
it is forbidden in the $\Xi^\prime,\Omega^\prime$ states,
and a heavy-quark must be exchanged to populate the lighter diquark sector from the $[cb]$.

\subsection{$3/2^+$ Spin-Flavour Combinations}
In the isospin-symmetry limit, using the quark-diquark approximation, the Faddeev equation in Fig.\,\ref{figFaddeev} supports the following $20$ distinct flavour-$SU(5)$ combinations for $3/2^+$ baryons.
{\allowdisplaybreaks
\begin{subequations}
\label{uDeltaColumn}
\begin{align}
u_\Delta & = \left[
\begin{array}{c}
\{uu\} u
\end{array} \right]
\leftrightarrow
\left[
\begin{array}{l}
{\mathpzc a}^{11}
\end{array}
\right], \label{B7a}\\
u_{\Sigma^\ast} & =
\left[ \begin{array}{c}
\{uu\} s \\
\{us\} u
\end{array} \right]
\leftrightarrow
\left[
\begin{array}{l}
{\mathpzc a}^{11}\\
{\mathpzc a}^{13}
\end{array}
\right], \label{B7b} \\
u_{\Xi^\ast} & =
\left[
\begin{array}{c}
\{us\} s \\
\{ss\} u
\end{array}\right]
\leftrightarrow
\left[
\begin{array}{l}
{\mathpzc a}^{13}\\
{\mathpzc a}^{20}
\end{array}
\right],\label{B7c}\\
u_\Omega & = \left[
\begin{array}{c}
\{ss\} s
\end{array} \right]
\leftrightarrow
\left[
\begin{array}{l}
{\mathpzc a}^{20}
\end{array}
\right], \label{B7d}\\
u_{\Sigma_{c}^{*++}} & =
\left[ \begin{array}{c}
\{uu\} c \\
\{uc\} u
\end{array} \right]
\leftrightarrow
\left[
\begin{array}{l}
{\mathpzc a}^{11}\\
{\mathpzc a}^{14}
\end{array}
\right],\label{B7e}\\
u_{\Xi_{c}^{*+}} & = \frac{1}{\sqrt{2}}
\left[
\begin{array}{c}
\sqrt{2} \{us\} c \\
\{uc\} s + \{sc\}u
\end{array}
\right]
\leftrightarrow
\left[
\begin{array}{l}
{\mathpzc a}^{13}\\
{\mathpzc a}^{\{14,21\}}
\end{array}
\right],\label{B7g}\\
u_{\Omega_{c}^{*0}} & =
\left[
\begin{array}{c}
\{ss\} c \\
\{sc\} s
\end{array}\right]
\leftrightarrow
\left[
\begin{array}{l}
{\mathpzc a}^{20}\\
{\mathpzc a}^{21}
\end{array}
\right],\label{B7f}\\
u_{\Sigma_{b}^{*+}} & =
\left[ \begin{array}{c}
\{uu\} b \\
\{ub\} u
\end{array} \right]
\leftrightarrow
\left[
\begin{array}{l}
{\mathpzc a}^{11}\\
{\mathpzc a}^{15}
\end{array}
\right],\label{B7h}\\
u_{\Xi_{b}^{*0}} & = \frac{1}{\sqrt{2}}
\left[
\begin{array}{c}
\sqrt{2} \{us\} b \\
\{ub\} s + \{sb\}u
\end{array}
\right]
\leftrightarrow
\left[
\begin{array}{l}
{\mathpzc a}^{13}\\
{\mathpzc a}^{\{15,22\}}
\end{array}
\right],\label{B7j}\\
u_{\Omega_{b}^{*-}} & =
\left[
\begin{array}{c}
\{ss\} b \\
\{sb\} s
\end{array}\right]
\leftrightarrow
\left[
\begin{array}{l}
{\mathpzc a}^{20}\\
{\mathpzc a}^{22}
\end{array}
\right],\label{B7i}\\
u_{\Xi_{cc}^{*++}} & =
\left[ \begin{array}{c}
\{uc\} c \\
\{cc\} u
\end{array} \right]
\leftrightarrow
\left[
\begin{array}{l}
{\mathpzc a}^{15}\\
{\mathpzc a}^{25}
\end{array}
\right],\label{B7k}\\
u_{\Xi_{cb}^{*+}} & = \frac{1}{\sqrt{2}}
\left[
\begin{array}{c}
\{uc\}b\! +\!  \{ub\}c\\
\sqrt{2} \{cb\}u
\end{array}
\right]
\leftrightarrow
\left[
\begin{array}{l}
{\mathpzc a}^{\{14,15\}}\\
{\mathpzc a}^{24}
\end{array}
\right],\label{B7m}\\
u_{\Xi_{bb}^{*0}} & =
\left[
\begin{array}{c}
\{ub\} b \\
\{bb\} u
\end{array}\right]
\leftrightarrow
\left[
\begin{array}{l}
{\mathpzc a}^{14}\\
{\mathpzc a}^{23}
\end{array}
\right],\label{B7l}\\
u_{\Omega_{cc}^{*+}} & =
\left[ \begin{array}{c}
\{sc\} c \\
\{cc\} s
\end{array} \right]
\leftrightarrow
\left[
\begin{array}{l}
{\mathpzc a}^{21}\\
{\mathpzc a}^{23}
\end{array}
\right],\label{B7n}\\
u_{\Omega_{cb}^{*0}} & = \frac{1}{\sqrt{2}}
\left[
\begin{array}{c}
\{sc\}b\! +\!  \{sb\}c\\
\sqrt{2} \{cb\}s
\end{array}
\right]
\leftrightarrow
\left[
\begin{array}{l}
{\mathpzc a}^{\{21,22\}}\\
{\mathpzc a}^{24}
\end{array}
\right],\label{B7p}\\
u_{\Omega_{bb}^{*-}} & =
\left[
\begin{array}{c}
\{sb\} b \\
\{bb\} s
\end{array}\right]
\leftrightarrow
\left[
\begin{array}{l}
{\mathpzc a}^{22}\\
{\mathpzc a}^{25}
\end{array}
\right],\label{B7o}\\
u_{\Omega_{ccc}^{\ast ++}} & = \left[
\begin{array}{c}
\{cc\} c
\end{array} \right]
\leftrightarrow
\left[
\begin{array}{l}
{\mathpzc a}^{23}
\end{array}
\right], \label{B7q}\\
u_{\Omega_{ccb}^{\ast+}} & =
\left[ \begin{array}{c}
\{cc\} b \\
\{cb\} c
\end{array} \right]
\leftrightarrow
\left[
\begin{array}{l}
{\mathpzc a}^{23}\\
{\mathpzc a}^{24}
\end{array}
\right], \label{B7r}\\
u_{\Omega_{cbb}^{\ast 0}} & =
\left[
\begin{array}{c}
\{cb\} b \\
\{bb\} c
\end{array}\right]
\leftrightarrow
\left[
\begin{array}{l}
{\mathpzc a}^{24}\\
{\mathpzc a}^{25}
\end{array}
\right],\label{B7s}\\
u_{\Omega_{bbb}^{\ast -}} & = \left[
\begin{array}{c}
\{bb\} b
\end{array} \right]
\leftrightarrow
\left[
\begin{array}{l}
{\mathpzc a}^{25}
\end{array}
\right].\label{B7t}
\end{align}
\end{subequations}
Evidently, only $J^P=1^+$ appear in these flavour-$SU(5)$ generalisations of the baryon decuplet.

\section{Selected Faddeev Equations}
\label{AppFEexplicit}
\subsection{$\Xi_c^+$ Baryon}
\label{FEexplicitXic}
With the $c$-quark as a label, there are two $SU(3)$ subgroups of a mixed-symmetric-$20$ representation of flavour-$SU(4)$: one is antisymmetric under the interchange of the lighter quarks (antitriplet); and the other is symmetric (sextet).  Both the antitriplet and sextet flavour combinations can be realised in the $J^P=1/2^+$ sector: the $\Xi_c^+$ is the antitriplet member, antisymmetric under $u\leftrightarrow s$, as expressed via Eq.\,\eqref{Xic}.  Using this knowledge, one can employ a procedure similar to that used for the $\Lambda$-baryon in Ref.\,\cite{Chen:2012qr}, and proceed from Fig.\,\ref{figFaddeev} to the following algebraic equation:
\begin{align}
\label{XicstarFaddeev}
u_{\Xi_c^+} & = {\mathpzc K}_{\,\Xi_c^{+}} \; u_{\Xi_c^{+}}\,,
\end{align}
where $u_{\Xi_c^{+}}$ is given in Eq.\,\eqref{Xic} and
\begin{align}
\label{XicFaddeevK}
{\mathpzc K}_{\,\Xi_c^{+}} & =
\left[
\begin{array}{ccc}
 0 & \displaystyle\frac{{\cal {\cal K}}_{2}^{3}+{\cal K}_{2}^{8}}{\sqrt{2}} &\displaystyle
   -\frac{{\cal K}_{2}^{14}+{\cal K}_{2}^{21}}{\sqrt{2}} \\
 \displaystyle\frac{{\cal K}_{3}^{2}+{\cal K}_{8}^{2}}{\sqrt{2}} &\displaystyle
   -\frac{{\cal K}_{3}^{8}+{\cal K}_{8}^{3}}{2} & \displaystyle-\frac{{\cal K}_{3}^{21}+{\cal K}_{8}^{14}}{2}
   \\
 \displaystyle-\frac{{\cal K}_{14}^{2}+{\cal K}_{21}^{2}}{\sqrt{2}} &
   \displaystyle-\frac{{\cal K}_{14}^{8}+{\cal K}_{21}^{3}}{2} &
  \displaystyle -\frac{{\cal K}_{14}^{21}+{\cal K}_{21}^{14}}{2} \\
\end{array}
\right]
\end{align}
with, \emph{e.g}.\
\begin{subequations}
\label{KernelIllustrate}
\begin{align}
\mathcal{K}_{2}^{3}&= -4\int\frac{d^4q}{(2\pi)^4}\,\Gamma_{3}(l)S_{u}^{\rm T} \bar\Gamma_{2}(-k)S_{s}(q)\Delta_{3}(l) \,,\\
\mathcal{K}_{3}^{2} &= -4\int\frac{d^4q}{(2\pi)^4}\,\Gamma_{2}(l)S_{u}^{\rm T}\bar\Gamma_{3}(-k)S_{c}(q)\Delta_{2}(l)\, .
\end{align}
\end{subequations}
The other elements of ${\mathpzc K}_{\,\Xi_c^{+}}$ have similar structures, determined by the Faddeev kernel in Fig.\,\ref{figFaddeev}, where each amplitude and propagator is connected with a particular diquark or quark according to the definitions in Appendix~\ref{AppFE}, \emph{e.g}.: $\Gamma_2$ is the correlation amplitude for the $[us]$ diquark;
$\Gamma_3$  the amplitude for the $[uc]$ diquark, with $\Delta_3$ the associated propagator;
and $S_{u,s}$ are propagators for the quarks with flavour $f=u,s$.

It is important to observe that $[{\mathpzc K}_{\,\Xi_c^{+}} ]_{11} \equiv 0$.   Hence, the $[us]c$ component of the $\Xi_c^{+}$ spin-flavour wave function is not ``self-supporting''.  Instead, it must be ``fed'' by other combinations.  This is a general consequence of Fig.\,\ref{figFaddeev}, \emph{i.e}.\ the outcome is not specific to the contact interaction.

Taking account of Eq.\,\eqref{calAcontact}, \emph{viz}.\ axial-vector diquarks have two Dirac structures, Eq.\,\eqref{XicFaddeevK} expands to a $4\times 4$ matrix because, \emph{e.g}.\
\begin{align}
{\cal K}^{14}_{2} = ({\cal K}_{2:1}^{14} , {\cal K}_{2:2}^{14})\,.
\end{align}

Now using the regularisation procedure described in Sec.\,\ref{SecIS}, all elements in the kernel matrix can be computed explicitly.  Defining
\begin{equation}
c_{\Xi_{c}}^f = \frac{g_{\Xi_{c}}^2}{4 \pi^2 M_f}\,,\;
\sigma_{\Xi_{c}}^{f,i} =
\sigma(\alpha,M_f^2,m_i^2,m_{\Xi_{c}}^2)\,,
%
\end{equation}
where $g_{\Xi_{c}}=g_2$ in Eq.\,\eqref{valueg2g4},
\begin{align}
\nonumber
\sigma(&\alpha,M_f^2,m_i^2,M_{\Xi_c}^2) \\
& = (1-\alpha)\,M_f^2 + \alpha\,m_i^2 - \alpha (1-\alpha) M_{\Xi_c}^2\,,\label{definesigma}
\end{align}
with $f$ being a quark flavour label and $i$ a diquark enumeration label, so that $m_i$ is the mass of the associated correlation, then the first row is:
{\allowdisplaybreaks
\begin{subequations}
\label{XicKernelReg}
\begin{align}
{\cal K}_{2}^{3}  = &
\frac{c_{\Xi_{c}}^u}{4M_{R_{2}}M_{R_{3}}} \!
\int_0^1 d\alpha \,\overline{\cal C}_1(\sigma_{\Xi_{c}}^{s,3})\nonumber \\
& \times [2 E_2 M_{R_{2}} - F_2 M_{\Xi_{c}} (1-\alpha)]\nonumber\\
& \times [2 E_3 M_{R_{3}} - F_3 M_{\Xi_{c}} (1-\alpha)] \nonumber\\
& \times  [M_s+\alpha M_{\Xi_{c}}] \,, \\
{\cal K}_{2}^{8}  = &
\frac{c_{\Xi_{c}}^s}{4M_{R_{2}}M_{R_{8}}} \!
\int_0^1 d\alpha \,\overline{\cal C}_1(\sigma_{\Xi_{c}}^{u,8})\nonumber \\
& \times [2 E_2 M_{R_{2}} - F_2 M_{\Xi_{c}} (1-\alpha)]\nonumber \\
& \times [2 E_8 M_{R_{8}} - F_8 M_{\Xi_{c}} (1-\alpha)]\nonumber\\
& \times  [M_u+\alpha M_{\Xi_{c}}]\,,\\
{\cal K}_{2:1}^{14}  = &
\frac{c_{\Xi_{c}}^u E_{14}}{2 M_{R_{2}} m_{14}^2} \!
\int_0^1 d\alpha \,\overline{\cal C}_1(\sigma_{\Xi_{c}}^{s,14})
\bigg[
2 E_2 M_{R_{2}} (3M_s m_{14}^2 \nonumber \\
& +M_{\Xi_{c}}[m_{14}^2+2M_{\Xi_{c}}^2(1-\alpha)^2]\alpha) \nonumber \\
& - F_2 M_{\Xi_{c}} (1-\alpha)[M_s(m_{14}^2 \nonumber \\
& +2M_{\Xi_{c}}^2 (1-\alpha)^2)+3m_{14}^2 M_{\Xi_{c}} \alpha]\bigg]\,,\\
{\cal K}_{2:2}^{14}  = & \frac{c_{\Xi_{c}}^u E_{14}}{2 M_{R_{2}} m_{14}^2} \!
\int_0^1 d\alpha \,\overline{\cal C}_1(\sigma_{\Xi_{c}}^{s,14})
[2 E_2 M_{R_{2}} \nonumber \\
& + F_2 M_{\Xi_{c}} (1-\alpha)]\nonumber \\
& \times [m_{14}^2 - M_{\Xi_{c}}^2(1-\alpha)^2][M_s-\alpha M_{\Xi_{c}}]\,, \\
{\cal K}_{2:1}^{21}  = & \frac{c_{\Xi_{c}}^s E_{21}}{2 M_{R_{2}} m_{21}^2} \!
\int_0^1 d\alpha \,\overline{\cal C}_1(\sigma_{\Xi_{c}}^{u,21})
\bigg[
2 E_2 M_{R_{2}} (3M_u m_{21}^2\nonumber\\
& +M_{\Xi_{c}}[m_{21}^2+2M_{\Xi_{c}}^2(1-\alpha)^2]\alpha)\nonumber \\
& - F_2 M_{\Xi_{c}} (1-\alpha)[M_u(m_{21}^2 \nonumber \\
& +2M_{\Xi_{c}}^2 (1-\alpha)^2)+3m_{21}^2 M_{\Xi_{c}} \alpha]\bigg]\,,\\
{\cal K}_{2:2}^{21}  = & \frac{c_{\Xi_{c}}^u E_{14}}{2 M_{R_{2}} m_{14}^2} \!
\int_0^1 d\alpha \,\overline{\cal C}_1(\sigma_{\Xi_{c}}^{s,14})
[2 E_2 M_{R_{2}} \nonumber \\
& + F_2 M_{\Xi_{c}} (1-\alpha)]\nonumber \\
& \times [m_{14}^2 - M_{\Xi_{c}}^2(1-\alpha)^2][M_s-\alpha M_{\Xi_{c}}]\,,
\end{align}
\end{subequations}
}
\hspace*{-0.4\parindent}%
where, \emph{e.g}.\ $M_{R_2} = M_u M_s/(M_u+M_s)$, because ``$2$'' labels the $[us]$ diquark, $E_2, F_2$ are the $[us]$-diquark amplitudes in Table~\ref{diquarkspectrum}, Row~2.  All other symbols are understood analogously.

The other elements of ${\mathpzc K}_{\,\Xi_c^{+}}$ are all defined by expressions like Eqs.\,\eqref{KernelIllustrate} and straightforward application of our regularisation procedure completes the array with expression like Eqs.\,\eqref{XicKernelReg}.

\subsection{$\Xi_c^{\prime+}$  Baryon}
$\Xi_c^{\prime +}$ is the sextet partner of the $\Xi_c^+$ in the mixed-symmetric-$20$, \emph{i.e}.\ the  $\Xi_c^{\prime +}$ spin-flavour wave function is symmetric under $u\leftrightarrow s$, \emph{viz}.\ Eq.\,\eqref{B7g}.  In this case, Fig.\,\ref{figFaddeev} yields the following equation:
\begin{align}
\label{XicprimeFaddeev}
u_{\Xi_c^{\prime +}} & = {\mathpzc K}_{\,\Xi_c^{\prime +}} \; u_{\Xi_c^{\prime +}}\,,
\end{align}
where $u_{\Xi_c^{\prime +}}$ is given in Eq.\,\eqref{Xicprime} and
\begin{align}
{\mathpzc K}_{\,\Xi_c^{\prime +}} & =
\left[
\begin{array}{ccc}
 \displaystyle \frac{{\cal K}_{3}^{8}+{\cal K}_{8}^{3}}{2}
   &\displaystyle
   -\frac{{\cal K}_{3}^{13}+{\cal K}_{8}^{13}}{\sqrt{2}}
   & \displaystyle\frac{{\cal K}_{3}^{21}+{\cal K}_{8}^{14}}{2} \\
 \displaystyle-\frac{{\cal K}_{13}^{3}+{\cal K}_{13}^{8}}{\sqrt{2}}
   & \displaystyle0
   &\displaystyle
   \frac{{\cal K}_{13}^{14}+{\cal K}_{13}^{21}}{\sqrt{2}} \\
 \displaystyle\frac{{\cal K}_{14}^{8}+{\cal K}_{21}^{3}}{2}
   & \displaystyle\frac{{\cal K}_{14}^{13}+{\cal K}_{21}^{13}}{\sqrt{2}}
   & \displaystyle\frac{{\cal K}_{14}^{21}+{\cal K}_{21}^{14}}{2}
\end{array}
\right]
\end{align}
Explicit forms for the entries in ${\mathpzc K}_{\,\Xi_c^{\prime +}} $ are readily obtained using the procedures indicated above; and they have forms similar to those in Eqs.\,\eqref{XicKernelReg}.

Here, since $[{\mathpzc K}_{\,\Xi_c^{\prime +}} ]_{22} \equiv 0$, the $\{us\}c$ component of the $\Xi_c^{\prime +}$ spin-flavour wave function is not ``self-supporting''.  On the other hand, it is the lightest possible diquark component of the bound-state and all feeder processes involve light-quark exchange.  Since the opposite is true for the other two correlations, \emph{viz}.\ $[uc]s+[sc]u$, $\{uc\}s+\{sc\}u$, then $\{us\}c$ can dominate in the $\Xi_c^{\prime +}$.  The contrast between this outcome for the $\Xi_c^{\prime +}$ and that described above for the $\Xi_c^+$ may be attributed to the different $u\leftrightarrow s$ symmetries of the systems, which can change constructive into destructive interference.

\subsection{$\Xi_{c}^{*}$ Baryon}
\label{AppXicstar}
$\Xi_{c}^{*}$ is a member of the symmetric-$20$ representation of flavour-$SU(4)$, in consequence of which its spin-flavour wave function must be symmetric under $u\leftrightarrow s$, \emph{viz}.\ Eq.\,\eqref{B7g}.
%
%
In this case, Fig.\,\ref{figFaddeev} yields an algebraic equation of the following form:
\begin{align}
\label{XicstarFaddeev}
u_{\Xi_c^{\ast +}} & = {\mathpzc K}_{\,\Xi_c^{\ast +}} \; u_{\Xi_c^{\ast +}}\,,
\end{align}
where
\begin{align}
\label{XicstarFaddeevK}
{\mathpzc K}_{\,\Xi_c^{\ast +}} & =
\left[
\begin{array}{cc}
0 & \displaystyle \frac{{\mathpzc K}_{13}^{14}+{\mathpzc K}_{13}^{21}}{\sqrt{2}}\\[2.5ex]
 \displaystyle\frac{{\mathpzc K}_{14}^{13}+{\mathpzc K}_{21}^{13}}{\sqrt{2}} &
 \displaystyle \frac{{\mathpzc K}_{14}^{21}+{\mathpzc K}_{21}^{14}}{2}
\end{array}\right]\,.
\end{align}
As usual, explicit expressions for the entries in $\mathcal{K}$ can be obtained using the procedures indicated above and they have forms similar to those in Eqs.\,\eqref{XicKernelReg}.

Evidently, $[\mathcal{K}]_{11} \equiv 0$ for the $\Xi_c^{\ast +}$.  Consequently, it is again the $\{us\}c$ component of the $\Xi_c^{\ast +}$ spin-flavour wave function that is not ``self-supporting''.  Instead, $\{us\}c$ feeds the compound $\{uc\}s+\{sc\}u$ correlation, which also supports itself and is therefore dominant.



\end{document}